\documentclass[10pt,a4paper]{article}
\usepackage{amsmath,amssymb,amsfonts}
\usepackage{graphicx}
\usepackage{hyperref}
\usepackage[margin=0.75in]{geometry}
\usepackage{float}
\usepackage{bm}
\usepackage{booktabs}
\usepackage{tabularx}

\usepackage{setspace}
\usepackage{titlesec}
\titlespacing*{\section}{0pt}{8pt}{4pt}
\titlespacing*{\subsection}{0pt}{6pt}{3pt}
\titlespacing*{\subsubsection}{0pt}{4pt}{2pt}

\AtBeginDocument{
  \setlength{\abovedisplayskip}{4pt}
  \setlength{\belowdisplayskip}{4pt}
  \setlength{\abovedisplayshortskip}{2pt}
  \setlength{\belowdisplayshortskip}{2pt}
}

\let\oldbibliography\thebibliography
\renewcommand{\thebibliography}[1]{%
  \oldbibliography{#1}%
  \setlength{\itemsep}{0pt}%
  \setlength{\parskip}{0pt}%
}

\title{\texttt{anyakrakusuma}: A Python Library for Entropic
Schr\"{o}dinger Bridges on Idealized Geometries}

\author{
Dasapta E. Irawan$^{1}$,
Sandy H. S. Herho$^{1,2,3,*}$,
Agus W. Jatmiko$^{4}$,\\
Sito F. Biosa$^{5}$,
Candrasa S. Dharma$^{6}$,
Edi Riawan$^{7}$,\\
Astyka Pamumpuni$^{1}$,
Rendy D. Kartiko$^{1}$,
Rusmawan Suwarman$^{7}$,\\
and Deny J. Puradimaja$^{1}$
}

\date{}

\def\Address{$^{1}$Applied Geology Research Group, Bandung Institute of
Technology, Bandung, West Java 40132, Indonesia\\
$^{2}$Department of Earth and Planetary Sciences, University of
California, Riverside, CA 92521, USA\\
$^{3}$Center for Agrarian Studies, Bandung Institute of Technology,
Bandung, West Java 40132, Indonesia\\
$^{4}$Headquarters of the Indonesian Air Force (Mabes TNI AU),
Cilangkap, East Jakarta 13870, Indonesia\\
$^{5}$Department of Visual Communication Design -- Animation, BINUS
University, West Jakarta 11480, Indonesia\\
$^{6}$Indonesian Navy Hydro-Oceanography Center (Pushidrosal), Ancol
Timur, North Jakarta 14430, Indonesia\\
$^{7}$Atmospheric Science Research Group, Bandung Institute of
Technology, Bandung, West Java 40132, Indonesia}

\def\corrAuthor{Corresponding Author}
\def\corrEmail{sh001@ucr.edu}

\begin{document}
\maketitle

\begin{center}
\small
\Address\\[2pt]
$^{*}$\corrAuthor: \href{mailto:\corrEmail}{\corrEmail}
\end{center}

\begin{abstract}
\noindent
We present \texttt{anyakrakusuma}, an open-source Python library
that solves the discrete static Schr\"{o}dinger bridge problem, the
entropically regularized counterpart of optimal transport, through a
log-domain Sinkhorn--Knopp iteration and reconstructs the entropic
interpolation between two empirical point clouds. The solver is paired with a
diagnostic pipeline that characterizes the optimal coupling and the
intermediate distributions through information-theoretic and geometric
measures. We exercise the library on four idealized planar cases spanning a
circle-to-circle dilation, a spiral-to-mixture fragmentation, a rigid
reorientation of two moons, and a Lissajous-to-trefoil deformation. The
log-domain formulation is necessary rather than merely convenient at the
parameters studied, where the cost-to-regularization ratio reaches four
hundred and the Gibbs kernel underflows double precision across most of its
range; the iteration nonetheless attains a marginal residual of $10^{-9}$ and
unit marginal fidelity in every case. Residual histories decay geometrically
over approximately eight decades at per-iteration contraction factors between
$0.966$ and $0.976$, which are local rates near the fixed point that lie many
orders of magnitude below the worst-case Hilbert-metric bound. The covariance
analysis recovers an imposed ninety-degree reorientation to within
$0.07^\circ$, roughly forty times smaller than its uncertainty, across a
masked interval of near-isotropy on which the principal axis is unobservable.
The diagnostics are reported with explicit attention to the regimes in which each is well defined, including the differential entropy, which is meaningful only on the open interpolation interval. The presented cases are constructed rather than measured; quantitative application to empirical point clouds requires further study.
\end{abstract}

\noindent\textbf{Keywords:}
entropic optimal transport; information-theoretic diagnostics;
log-domain Sinkhorn algorithm; Python scientific software;
Schr\"{o}dinger bridge problem.

\section{Introduction}

The Schr\"{o}dinger bridge problem asks for the most likely evolution of a
cloud of particles between two observed distributions, given a reference
stochastic dynamics, and was posed in its original form as a question about
the statistical behavior of diffusing particles conditioned on their initial
and final states \cite{schrodinger1931,schrodinger1932}. In modern terms the
problem is an entropy-minimizing interpolation on path space, and it is now
understood to be the stochastic, entropically regularized counterpart of the
optimal transport problem, to which it reduces in the vanishing-noise limit
\cite{leonard2014,chen2016}. This correspondence has placed the Schr\"{o}dinger
bridge at the intersection of probability, statistical mechanics, and
computational mathematics, where it supplies a principled way to connect two
empirical states through a dynamics that is neither purely deterministic nor
purely diffusive \cite{follmer1988,villani2009}.

Interest in the problem has grown with the recognition that its discrete
static form is solved by the same matrix-scaling iteration that underlies
entropic optimal transport. The introduction of an entropic penalty renders
the transport problem strictly convex and solvable by the Sinkhorn--Knopp
iteration at a cost far below that of the underlying linear program
\cite{cuturi2013,sinkhornknopp1967}, and the same machinery, together with its
Bregman-projection and stabilized-scaling refinements, now forms the standard
computational backbone for entropic transport
\cite{peyrecuturi2019,benamou2015,schmitzer2019}. The resulting framework has
been carried well beyond its origins, from the continuous-state generative
models that treat the bridge as the finite-time analogue of a diffusion
\cite{debortoli2021} to applications across the physical and data sciences
where two snapshots of a system must be joined by a plausible intervening
process. The breadth of this adoption has made the quality and transparency of
the underlying solvers a matter of practical consequence.

That breadth has not been matched by a corresponding supply of reusable,
well-documented scientific software. Implementations of the entropic bridge
are frequently embedded within larger machine-learning pipelines, tuned to a
single application, and released without the archival data formats,
reproducible configurations, or standardized diagnostics that characterize
mature community codes in adjacent computational fields. A researcher who
wishes to study the bridge itself, rather than a downstream task built upon
it, is often left to reconstruct the solver and its analysis from scratch. The
absence of a purpose-built, openly documented tool for the discrete
Schr\"{o}dinger bridge, coupled with information-theoretic diagnostics and
climate-and-forecast-compliant output, impedes systematic exploration,
cross-study comparison, and reproducibility in the same way that has been
noted for other idealized-modeling communities \cite{harris2020,lam2015}.

A productive response to this situation has been to build compact,
single-purpose solvers whose numerical core is legible and just-in-time
compiled for research-grade throughput, and to pair each with a diagnostic
layer that quantifies the organization of the solution beyond first-moment
summaries and with self-describing output that supports archival and reuse.
Idealized solvers constructed in this manner have proven effective across a
range of physical settings, including shear-driven instability
\cite{herho2025kh2d}, collective animal motion
\cite{herho2026dewikadita}, nonlinear dispersive waves
\cite{irawan2026sangkuriang}, and wave attenuation in coastal environments
\cite{herho2026waveatten}, and the same emphasis on numerical transparency
underlies reproducibility studies across computing platforms
\cite{herho2025pendulum,herho2025optionmc}. Information-theoretic diagnostics
in particular have repeatedly exposed organizational structure that
order-parameter or first-moment descriptions leave hidden
\cite{herho2026dewikadita,irawan2026sangkuriang}. The entropic Schr\"{o}dinger
bridge is a natural setting for the same treatment, since its coupling and its
intermediate distributions carry precisely the kind of structure that such
diagnostics are designed to measure.

This paper presents \texttt{anyakrakusuma}, an open-source \texttt{Python}
library that solves the discrete static Schr\"{o}dinger bridge problem through
a log-domain Sinkhorn--Knopp iteration and reconstructs the entropic
interpolation between two empirical point clouds, together with a diagnostic
pipeline that characterizes the coupling and the intermediate distributions
through information-theoretic and geometric measures. The solver and its
diagnostics are exercised on four idealized planar cases, chosen to span a
circle-to-circle dilation, a spiral-to-mixture fragmentation, a rigid
reorientation of two moons, and a Lissajous-to-trefoil deformation, which
between them probe convergence, the structure of the optimal coupling, the
evolution of the intermediate density, and the informational and geometric
descriptors of the bridge. Consistent with the aim of providing a verified
computational foundation rather than an application to measured data, the four
cases are constructed rather than observed, and the diagnostics are reported
with explicit attention to the regimes in which each is well defined. The
mathematical formulation and numerical implementation are developed first,
followed by the four demonstration cases, their diagnostic analysis, and a
discussion of the results and their limitations.
\section{Methods}
\subsection{Model Description}

The problem treated in this work was first formulated by Erwin Schr\"odinger
in two papers in the early 1930s~\cite{schrodinger1931,schrodinger1932}.
Consider an ensemble of independent Brownian particles whose spatial
distribution is recorded at two distinct instants in time, and suppose that
the recorded distributions are mutually inconsistent with free diffusion.
Schr\"odinger's problem is to identify the most likely evolution of the
ensemble between the two observations. Modern probabilistic and analytical
treatments have made the question precise and have shown that its resolution
coincides with the solution of an entropy-regularized version of the
Monge--Kantorovich optimal transport
problem~\cite{leonard2014,chen2016,peyrecuturi2019}. The following
development establishes the correspondence in three stages. A path-space
formulation grounded in the relative entropy of measures reduces to a
coupling problem on the product of the endpoint spaces, and the resulting
variational problem is then cast in the finite-dimensional form that governs
the discrete computation.

Let $d \in \mathbb{N}$ denote the spatial dimension and $\varepsilon > 0$ a
fixed positive parameter. Consider independent particles diffusing in
$\mathbb{R}^d$ on the unit time interval $t \in [0,1]$ according to the
stochastic differential equation
\begin{equation}
\mathrm{d}X_t \;=\; \sqrt{\varepsilon}\, \mathrm{d}W_t,
\qquad X_0 \,\sim\, \rho_0,
\label{eq:reference_sde}
\end{equation}
where $X_t \in \mathbb{R}^d$ is the particle position at time $t$, $W_t$ is
a standard $d$-dimensional Wiener process, and $\rho_0 : \mathbb{R}^d \to
\mathbb{R}_{\geq 0}$ is the prescribed probability density of the initial
position. The parameter $\varepsilon$ plays the role of a diffusivity.
Throughout what follows, $\Omega := C([0,1]; \mathbb{R}^d)$ denotes the
space of continuous paths $\omega : [0,1] \to \mathbb{R}^d$, equipped with
the supremum norm, and $\mathcal{P}(\Omega)$ denotes the Borel probability
measures on $\Omega$. The reference path measure $R \in \mathcal{P}(\Omega)$
is the law of the diffusion~\eqref{eq:reference_sde}, that is, the joint
distribution of the random trajectory $\{X_t\}_{t \in [0,1]}$. For a path
$\omega \in \Omega$ and a time $t \in [0,1]$, $\omega_t \in \mathbb{R}^d$
denotes the position at time $t$.

Suppose that at the terminal time $t = 1$ a second density $\rho_1 :
\mathbb{R}^d \to \mathbb{R}_{\geq 0}$ is observed, and that $\rho_1$ does not
match the one-time marginal of $X_1$ predicted by~\eqref{eq:reference_sde}.
Schr\"odinger's problem selects the path measure $P^\star$ closest to $R$ in
relative entropy among those whose endpoint marginals are exactly $(\rho_0,
\rho_1)$~\cite{leonard2014,chen2016},
\begin{equation}
P^\star \;=\; \mathop{\mathrm{arg\,min}}_{P \,\in\, \Pi(\rho_0, \rho_1)}
\;\mathrm{KL}\!\left( P \,\|\, R \right).
\label{eq:dynamic_sbp}
\end{equation}
Here $\Pi(\rho_0, \rho_1) \subset \mathcal{P}(\Omega)$ is the set of path
measures whose endpoint pushforwards match the prescribed marginals,
\begin{equation}
\Pi(\rho_0, \rho_1) \;=\; \left\{
P \in \mathcal{P}(\Omega) \,:\,
P \circ \omega_0^{-1} = \rho_0,\;
P \circ \omega_1^{-1} = \rho_1
\right\},
\label{eq:path_polytope}
\end{equation}
and $\mathrm{KL}(P \,\|\, R)$ is the Kullback--Leibler (KL) divergence,
\begin{equation}
\mathrm{KL}\!\left( P \,\|\, R \right) \;=\;
\int_\Omega \log\!\left( \frac{\mathrm{d}P}{\mathrm{d}R}(\omega) \right)
\, \mathrm{d}P(\omega),
\label{eq:kl_definition}
\end{equation}
defined whenever $P$ is absolutely continuous with respect to $R$, with
$\mathrm{d}P/\mathrm{d}R$ the Radon--Nikodym derivative, and as $+\infty$
otherwise~\cite{kullback1951}. The minimizer $P^\star$, when it exists,
admits an interpretation through the theory of large deviations as the most
likely empirical distribution of a large collection of independent
trajectories of~\eqref{eq:reference_sde}, conditional on the empirical
endpoint marginals matching $(\rho_0, \rho_1)$~\cite{follmer1988}.

The problem stated in~\eqref{eq:dynamic_sbp} is posed on the
infinite-dimensional path space $\Omega$, and its direct discretization is
neither obvious nor efficient. A classical reduction lowers the problem to
one posed on the product of the endpoint spaces. Let $R_{0,1} \in
\mathcal{P}(\mathbb{R}^d \times \mathbb{R}^d)$ denote the joint law of the
endpoints $(\omega_0, \omega_1)$ under $R$. For the
diffusion~\eqref{eq:reference_sde}, the transition density is the Gaussian
heat kernel at time one, so $R_{0,1}$ admits the explicit factorization
\begin{equation}
R_{0,1}(\mathrm{d}x,\mathrm{d}y) \;=\;
\rho_0(x)\, k_\varepsilon(x,y)\, \mathrm{d}x\, \mathrm{d}y,
\qquad
k_\varepsilon(x,y) \;=\; (2\pi\varepsilon)^{-d/2}\,
\exp\!\left( -\frac{\|x - y\|^2}{2\varepsilon} \right),
\label{eq:reference_endpoint}
\end{equation}
where $x, y \in \mathbb{R}^d$ are the endpoint positions, $\|\cdot\|$
denotes the Euclidean norm on $\mathbb{R}^d$, and $k_\varepsilon$ is the
density of $X_1$ given $X_0 = x$ under the reference dynamics. Any path
measure $P \in \mathcal{P}(\Omega)$ admits a unique disintegration into its
endpoint marginal $\pi \in \mathcal{P}(\mathbb{R}^d \times \mathbb{R}^d)$,
defined by
\begin{equation}
\pi(\mathrm{d}x, \mathrm{d}y) \;=\;
P\!\left( \omega_0 \in \mathrm{d}x,\; \omega_1 \in \mathrm{d}y \right),
\label{eq:endpoint_marginal}
\end{equation}
together with the family of conditional path measures
\begin{equation}
P^{x,y}(\cdot) \;=\; P\!\left( \,\cdot\; \big|\; \omega_0 = x,\; \omega_1 = y
\right), \qquad (x,y) \in \mathbb{R}^d \times \mathbb{R}^d.
\label{eq:conditional_path}
\end{equation}
The analogous disintegration of $R$ produces the conditional family
$\{R^{x,y}\}$, in which $R^{x,y}$ is the law of the Brownian bridge of
diffusivity $\varepsilon$ on $[0,1]$ that starts at $x$ and ends at
$y$~\cite{follmer1988}. The chain rule for relative entropy then yields the
additive decomposition~\cite{leonard2014}
\begin{equation}
\mathrm{KL}\!\left( P \,\|\, R \right) \;=\;
\mathrm{KL}\!\left( \pi \,\|\, R_{0,1} \right) \;+\;
\int_{\mathbb{R}^d \times \mathbb{R}^d}
\mathrm{KL}\!\left( P^{x,y} \,\|\, R^{x,y} \right)\, \pi(\mathrm{d}x, \mathrm{d}y).
\label{eq:kl_decomposition}
\end{equation}
The two terms on the right of~\eqref{eq:kl_decomposition} are coupled only
through the marginal $\pi$, and the second term is minimized term by term
in $(x,y)$ by setting the conditional law to its reference value
$P^{\star,x,y} = R^{x,y}$. The minimization in~\eqref{eq:dynamic_sbp}
therefore reduces to a problem for the endpoint coupling alone,
\begin{equation}
\pi^\star \;=\; \mathop{\mathrm{arg\,min}}_{\pi \,\in\, \Pi(\rho_0, \rho_1)}
\;\mathrm{KL}\!\left( \pi \,\|\, R_{0,1} \right),
\label{eq:static_sbp}
\end{equation}
where $\Pi(\rho_0, \rho_1)$ now denotes the set of probability measures on
the product space $\mathbb{R}^d \times \mathbb{R}^d$ whose first and second
marginals equal $\rho_0$ and $\rho_1$, respectively. The original dynamic
problem is recovered from $\pi^\star$ by gluing the Brownian bridge laws
$R^{x,y}$ along the optimal endpoint coupling.

Substituting the explicit form of $R_{0,1}$
from~\eqref{eq:reference_endpoint} into the relative entropy
in~\eqref{eq:static_sbp} and discarding constants that do not depend on
$\pi$ produces
\begin{equation}
\pi^\star \;=\; \mathop{\mathrm{arg\,min}}_{\pi \,\in\, \Pi(\rho_0, \rho_1)}
\;\int_{\mathbb{R}^d \times \mathbb{R}^d} \frac{\|x - y\|^2}{2}\,
\mathrm{d}\pi(x,y)
\;+\; \varepsilon\, \mathrm{KL}\!\left( \pi \,\|\, \rho_0 \otimes \rho_1 \right),
\label{eq:eot_continuous}
\end{equation}
where $\rho_0 \otimes \rho_1$ denotes the independent product measure on
$\mathbb{R}^d \times \mathbb{R}^d$. The functional
in~\eqref{eq:eot_continuous} is the Monge--Kantorovich optimal transport
cost with quadratic ground cost, regularized by an entropic term of
strength $\varepsilon$~\cite{brenier1991,villani2009}. As $\varepsilon \to
0^+$, the minimizer of~\eqref{eq:eot_continuous} converges to the
unregularized Wasserstein-optimal coupling, which is supported on the graph
of the Brenier map~\cite{brenier1991,mikami2004,carlier2017}. In the
opposite limit $\varepsilon \to \infty$, the entropic term dominates the
cost and the minimizer converges to the product measure $\rho_0 \otimes
\rho_1$, in which the source and target are statistically independent.

For empirical measures supported on finite point clouds, the
problem~\eqref{eq:eot_continuous} becomes finite-dimensional. Let
\begin{equation}
\mu \;=\; \sum_{i=1}^n a_i\, \delta_{x_i},
\qquad
\nu \;=\; \sum_{j=1}^m b_j\, \delta_{y_j},
\label{eq:empirical_marginals}
\end{equation}
where $\{x_i\}_{i=1}^n$ and $\{y_j\}_{j=1}^m$ are points in $\mathbb{R}^d$,
$\delta_x$ is the Dirac mass at $x$, and the weight vectors $a \in
\mathbb{R}^n_{> 0}$ and $b \in \mathbb{R}^m_{> 0}$ lie on the open
probability simplices, that is, $\sum_i a_i = \sum_j b_j = 1$ with all
entries positive. A coupling between $\mu$ and $\nu$ is represented by a
nonnegative matrix $\pi \in \mathbb{R}^{n \times m}_{\geq 0}$ with entries
$\pi_{ij} \geq 0$, $i = 1, \ldots, n$, $j = 1, \ldots, m$, interpreted as
the mass transported from $x_i$ to $y_j$. Define the cost matrix $C \in
\mathbb{R}^{n \times m}_{\geq 0}$ by
\begin{equation}
C_{ij} \;=\; \|x_i - y_j\|^2,
\label{eq:cost_matrix}
\end{equation}
and the transportation polytope
\begin{equation}
\Pi(a, b) \;=\; \left\{
\pi \in \mathbb{R}^{n \times m}_{\geq 0} \,:\,
\pi\, \mathbf{1}_m = a,\;
\pi^\top\, \mathbf{1}_n = b
\right\},
\label{eq:transport_polytope}
\end{equation}
where $\mathbf{1}_k \in \mathbb{R}^k$ denotes the column vector of all
ones. The continuous problem~\eqref{eq:eot_continuous} then becomes the
discrete convex program
\begin{equation}
\pi^\star \;=\; \mathop{\mathrm{arg\,min}}_{\pi \,\in\, \Pi(a, b)}
\;\langle C, \pi \rangle_F
\;+\; \varepsilon\, \sum_{i=1}^n \sum_{j=1}^m
\pi_{ij}\, \log\!\left( \frac{\pi_{ij}}{a_i b_j} \right),
\label{eq:eot_discrete}
\end{equation}
where $\langle C, \pi \rangle_F = \sum_{ij} C_{ij}\, \pi_{ij}$ is the
Frobenius inner product of $C$ and $\pi$, and the convention $0 \log 0 = 0$
is used. The factor of one half that appears in the quadratic cost
of~\eqref{eq:eot_continuous} has been absorbed into a redefinition of
$\varepsilon$, following the convention standard in the computational
entropic transport literature~\cite{peyrecuturi2019}.

The convex program~\eqref{eq:eot_discrete} admits a closed-form
characterization of its minimizer. Introducing Lagrange multipliers $f \in
\mathbb{R}^n$ and $g \in \mathbb{R}^m$ for the row and column marginal
constraints in~\eqref{eq:transport_polytope}, and setting the gradient of
the Lagrangian with respect to $\pi_{ij}$ to zero, gives the first-order
optimality condition
\begin{equation}
\pi^\star_{ij} \;=\;
a_i\, b_j\, \exp\!\left( \frac{f_i + g_j - C_{ij}}{\varepsilon} \right)
\;=\; u_i\, K_{ij}\, v_j,
\qquad
i = 1, \ldots, n,\;
j = 1, \ldots, m,
\label{eq:gibbs_factor}
\end{equation}
where the scaling vectors $u \in \mathbb{R}^n_{> 0}$ and $v \in
\mathbb{R}^m_{> 0}$ are given by $u_i = a_i\, \exp(f_i / \varepsilon)$ and
$v_j = b_j\, \exp(g_j / \varepsilon)$, and the Gibbs kernel $K \in
\mathbb{R}^{n \times m}_{> 0}$ has entries
\begin{equation}
K_{ij} \;=\; \exp\!\left( -\frac{C_{ij}}{\varepsilon} \right).
\label{eq:gibbs_kernel}
\end{equation}
The matrix-scaling structure $\pi^\star = \mathrm{diag}(u)\, K\,
\mathrm{diag}(v)$ has been studied since the 1960s, and the pair $(u, v)$
is known to exist and to be unique up to the rescaling $(u, v) \mapsto
(\alpha u, \alpha^{-1} v)$ for $\alpha > 0$, given that the marginal
constraints are imposed~\cite{sinkhorn1964,sinkhornknopp1967}. The discrete
scalings $(u, v)$ are point-cloud realizations of a pair of positive
functions $\varphi, \psi : \mathbb{R}^d \times [0,1] \to \mathbb{R}_{> 0}$
that solve the coupled Schr\"odinger system,
\begin{align}
\partial_t \varphi(x, t) \;&=\; \frac{\varepsilon}{2}\, \Delta \varphi(x, t),
\label{eq:schrodinger_phi} \\
\partial_t \psi(x, t) \;&=\; -\frac{\varepsilon}{2}\, \Delta \psi(x, t),
\label{eq:schrodinger_psi}
\end{align}
where $\Delta = \sum_{k=1}^d \partial^2 / \partial x_k^2$ is the Laplacian
on $\mathbb{R}^d$. The intermediate-time density factors multiplicatively as
$\rho_t(x) = \varphi(x, t)\, \psi(x, t)$, and the boundary conditions
$\varphi(\cdot, 0)\, \psi(\cdot, 0) = \rho_0$ and $\varphi(\cdot, 1)\,
\psi(\cdot, 1) = \rho_1$ close the system at both
ends~\cite{leonard2014,chen2016}. Equations~\eqref{eq:schrodinger_phi}
and~\eqref{eq:schrodinger_psi} consist of a forward heat equation for
$\varphi$ and a formally time-reversed heat equation for $\psi$. The
time-reversed companion distinguishes Schr\"odinger's formulation from
standard parabolic theory and motivated his interest in the problem as an
analogue of quantum-mechanical probability conservation.

Reconstruction of the intermediate-time density $\rho_t$ for $t \in (0, 1)$
from the optimal coupling $\pi^\star$ exploits the conditional-bridge
identity established between~\eqref{eq:kl_decomposition}
and~\eqref{eq:static_sbp}, which restores Brownian bridge dynamics on each
pair of endpoints. Conditional on $(X_0, X_1) = (x, y)$ drawn from
$\pi^\star$, the optimal process $X_t$ is the Brownian bridge of diffusivity
$\varepsilon$ between $x$ and $y$ on $[0, 1]$, whose marginal at time $t$
is Gaussian,
\begin{equation}
X_t \,\big|\, (X_0 = x,\, X_1 = y)
\;\sim\;
\mathcal{N}\!\left(
(1 - t)\, x + t\, y,\; \varepsilon\, t (1 - t)\, I_d
\right),
\label{eq:bridge_marginal}
\end{equation}
where $\mathcal{N}(\boldsymbol{\mu}, \boldsymbol{\Sigma})$ denotes the
multivariate normal distribution on $\mathbb{R}^d$ with mean vector
$\boldsymbol{\mu} \in \mathbb{R}^d$ and covariance matrix
$\boldsymbol{\Sigma} \in \mathbb{R}^{d \times d}$, and $I_d \in
\mathbb{R}^{d \times d}$ is the identity matrix~\cite{follmer1988}.
Integrating~\eqref{eq:bridge_marginal} against the optimal coupling yields
the intermediate-time density,
\begin{equation}
\rho_t(z) \;=\;
\int_{\mathbb{R}^d \times \mathbb{R}^d}
\mathcal{N}\!\left(
z;\; (1 - t)\, x + t\, y,\; \varepsilon\, t (1 - t)\, I_d
\right)\, \mathrm{d}\pi^\star(x, y),
\qquad z \in \mathbb{R}^d,
\label{eq:rho_t}
\end{equation}
where $\mathcal{N}(z; \boldsymbol{\mu}, \boldsymbol{\Sigma})$ is the value
of the Gaussian density at $z$. The variance $\varepsilon\, t (1 - t)$
along the bridge attains its maximum $\varepsilon / 4$ at $t = 1/2$ and
vanishes at the two endpoints $t = 0$ and $t = 1$, so the boundary
marginals $\rho_0$ and $\rho_1$ are recovered exactly. In the limit
$\varepsilon \to 0^+$, the Gaussian kernel in~\eqref{eq:rho_t} contracts to
a Dirac mass on the segment $(1 - t)\, x + t\, y$, and the intermediate
density reduces to the $L^2$-Wasserstein displacement interpolation
associated with the Brenier map~\cite{brenier1991}.

Equations~\eqref{eq:reference_sde}--\eqref{eq:rho_t} specify the model in
full. The discrete optimal coupling $\pi^\star$ defined
in~\eqref{eq:eot_discrete}, the dual scaling pair $(u, v)$ that produces it
through~\eqref{eq:gibbs_factor}, and the intermediate-time density $\rho_t$
reconstructed from~\eqref{eq:rho_t} are the three quantities that the
formulation delivers. The parameter $\varepsilon$ plays a double role,
serving simultaneously as the diffusivity of the reference Brownian motion
and as the strength of the entropic regularization
in~\eqref{eq:eot_discrete}, with small values producing sharp couplings
that approach the unregularized optimal transport plan and large values
producing diffuse couplings that approach statistical independence between
source and target.

\subsection{Numerical Implementation}

The routines that translate the discrete
formulation~\eqref{eq:eot_discrete}--\eqref{eq:rho_t} into executable
computations are packaged as the \texttt{anyakrakusuma} library, a Python
implementation in which all inner loops are accelerated by \texttt{Numba}
just-in-time compilation to LLVM intermediate representation and dispatched
across CPU threads through the \texttt{prange}
construct~\cite{lam2015}, while all array storage and higher-level
operations rely on \texttt{NumPy}~\cite{harris2020}. Numba compilation is
optional at import time and a pure \texttt{NumPy} fallback is provided for
environments in which the compiler is unavailable, at the cost of a
substantial slowdown on the inner Sinkhorn loops. All floating-point
arithmetic is performed in IEEE 754 double precision, and the just-in-time
decorators are configured with the \texttt{fastmath=True} option, which
permits associative reordering of floating-point additions to enable
vectorization at the price of strict bitwise reproducibility across compiler
versions~\cite{goldberg1991}. The associated numerical drift is not
consequential for the algorithm at hand, since the log-domain Sinkhorn
updates are themselves invariant under shifts of the potentials that far
exceed the accumulated fused-multiply-add error. The main entry point is
the \texttt{SchrodingerBridgeSolver} class, which exposes the point-cloud
transport problem through a \texttt{solve} method that returns the pair
$(f, g)$ together with the materialized plan and diagnostic quantities, and
a \texttt{generate\_trajectory} method that produces the bridge
interpolation at a uniform time grid.

The squared-Euclidean cost matrix $C \in \mathbb{R}^{n \times m}_{\geq 0}$
of~\eqref{eq:cost_matrix} is assembled through a triply nested loop over
the source index $i \in \{1, \ldots, n\}$, the target index $j \in \{1,
\ldots, m\}$, and the coordinate index $k \in \{1, \ldots, d\}$,
\begin{equation}
C_{ij} \;=\; \sum_{k=1}^d ( x_{ik} - y_{jk} )^2,
\qquad
i = 1, \ldots, n,\;
j = 1, \ldots, m,
\label{eq:cost_assembly}
\end{equation}
with the outer loop over $i$ parallelized through \texttt{prange}. The
inner two loops are kept explicit rather than replaced by vectorized
broadcasting because Numba compiles the explicit form to tight machine code
with cache-friendly access patterns on the row-major \texttt{NumPy} arrays,
avoiding the temporary allocations that a broadcasting expression would
generate. Storage for $C$ is preallocated as a contiguous $n \times m$
\texttt{float64} array, giving a memory footprint of $8 n m$ bytes and a
computational cost of $O(n m d)$ elementary operations for the assembly.

The Sinkhorn iterations that solve~\eqref{eq:eot_discrete} are executed in
the logarithmic domain, replacing the exponential-domain scalings $u_i$ and
$v_j$ of~\eqref{eq:gibbs_factor} by the dual potentials
\begin{equation}
f_i \;=\; \varepsilon\, \log u_i,
\qquad
g_j \;=\; \varepsilon\, \log v_j,
\qquad
i = 1, \ldots, n,\;
j = 1, \ldots, m.
\label{eq:log_potentials}
\end{equation}
Working with $(f, g)$ rather than $(u, v)$ removes the risk of underflow
or overflow when $\varepsilon$ is small relative to the range of $C_{ij}$,
a regime in which the Gibbs kernel entries $K_{ij} =
\exp(-C_{ij}/\varepsilon)$ of~\eqref{eq:gibbs_kernel} can lie many decades
below the smallest representable double-precision number. The log-domain
formulation and its stabilized variants have become standard practice for
entropic transport at small regularization, both in the balanced setting
treated here~\cite{schmitzer2019,peyrecuturi2019} and in the more general
scaling framework developed for unbalanced and multi-marginal
extensions~\cite{chizat2018,benamou2015}. The load-bearing primitive is
the numerically stable log-sum-exp (LSE) evaluated with the max-trick,
\begin{equation}
\mathrm{LSE}(z_1, \ldots, z_L) \;=\; z_{\max} \;+\;
\log \sum_{\ell = 1}^{L} \exp(z_\ell - z_{\max}),
\qquad
z_{\max} \;=\; \max_{\ell} z_\ell,
\label{eq:lse}
\end{equation}
implemented as a self-contained Numba-compiled routine with a special-case
branch that returns $z_{\max}$ directly whenever the maximum is infinite,
so that empty softmin evaluations do not propagate \texttt{NaN} through
the subsequent arithmetic.

In terms of $(f, g)$ and the KL optimality conditions
produced by~\eqref{eq:eot_discrete}, one full Sinkhorn sweep amounts to the
pair of updates
\begin{align}
f_i^{(k+1)} \;&=\; \varepsilon\, \log a_i \;-\;
\varepsilon\, \mathrm{LSE}_j\!\left( \frac{g_j^{(k)} - C_{ij}}{\varepsilon} \right),
\qquad i = 1, \ldots, n,
\label{eq:f_update} \\
g_j^{(k+1)} \;&=\; \varepsilon\, \log b_j \;-\;
\varepsilon\, \mathrm{LSE}_i\!\left( \frac{f_i^{(k+1)} - C_{ij}}{\varepsilon} \right),
\qquad j = 1, \ldots, m,
\label{eq:g_update}
\end{align}
in which the superscript denotes the iteration index and the second update
uses the freshly computed $f^{(k+1)}$ rather than the stale $f^{(k)}$,
corresponding to a Gauss--Seidel rather than a Jacobi sweep. A direct
consequence of the ordering in~\eqref{eq:f_update}--\eqref{eq:g_update} is
that the column-marginal constraint of the transportation
polytope~\eqref{eq:transport_polytope} is satisfied exactly at the end of
every sweep,
\begin{equation}
\sum_{i=1}^n \exp\!\left( \frac{f_i^{(k+1)} + g_j^{(k+1)} - C_{ij}}{\varepsilon} \right)
\;=\; b_j,
\qquad j = 1, \ldots, m,
\label{eq:col_marginal_exact}
\end{equation}
while the row marginal $\sum_j \pi_{ij}^{(k+1)}$ deviates from $a_i$ by an
amount that decreases monotonically as the iteration proceeds. The outer
loops in~\eqref{eq:f_update} and~\eqref{eq:g_update} are parallelized
across threads, while the inner LSE evaluation is kept serial per
thread to preserve the deterministic max-trick evaluation order, giving
each full sweep a computational cost of $O(n m)$ elementary operations
after the cost matrix has been assembled. The
formulation~\eqref{eq:f_update}--\eqref{eq:g_update} is the standard
log-domain Sinkhorn recursion~\cite{peyrecuturi2019,benamou2015}, and it
reduces to the classical multiplicative Sinkhorn--Knopp iteration on the
scalings $(u, v)$ in the limit of large
$\varepsilon$~\cite{sinkhorn1964,sinkhornknopp1967,knight2008}.

Both potentials are initialized to the zero vector, $f^{(0)} = 0$ and
$g^{(0)} = 0$, corresponding to the independent product coupling $u_i^{(0)}
v_j^{(0)} = a_i b_j$ in~\eqref{eq:gibbs_factor} and offering no informative
prior to bias the iteration toward any particular transport structure. The
logarithms of the marginal weight vectors, $\log a$ and $\log b$, are
precomputed once at solver entry after adding a floor of $10^{-300}$
inside the logarithm to guard against strictly zero entries in
user-supplied marginals; for the default uniform weights $a_i = 1/n$ and
$b_j = 1/m$ this floor is inactive. Convergence is monitored through the
$\ell^1$ residual of the row marginal constraint
in~\eqref{eq:transport_polytope},
\begin{equation}
\eta^{(k)} \;=\; \left\| \pi^{(k)} \mathbf{1}_m - a \right\|_1
\;=\;
\sum_{i=1}^n \left| \sum_{j=1}^m \pi_{ij}^{(k)} - a_i \right|,
\label{eq:marginal_residual}
\end{equation}
which is evaluated without materializing the plan through the log-domain
identity
\begin{equation}
\log\!\left( \pi^{(k)} \mathbf{1}_m \right)_i
\;=\;
\frac{f_i^{(k)}}{\varepsilon} \;+\;
\mathrm{LSE}_j\!\left( \frac{g_j^{(k)} - C_{ij}}{\varepsilon} \right),
\qquad i = 1, \ldots, n,
\label{eq:log_row_marginal}
\end{equation}
which is derived by taking the logarithm of the row sum of the Gibbs
factorization~\eqref{eq:gibbs_factor} and grouping the constant $f_i^{(k)}
/ \varepsilon$ outside the LSE. The residual~\eqref{eq:marginal_residual}
is evaluated at every tenth Sinkhorn sweep rather than at every sweep to
amortize the associated $O(nm)$ cost across the intervening iterations.
The check is deliberately asymmetric, using only the row-marginal residual
rather than the sum of row and column residuals, since the column-marginal
identity~\eqref{eq:col_marginal_exact} already holds up to floating-point
precision at the end of each sweep. The convergence properties of the
Sinkhorn recursion under mild positivity assumptions on the marginals were
established in the classical setting~\cite{sinkhorn1964,knight2008}, and
the log-domain adaptation preserves the geometric contraction rate in the
Hilbert projective metric that underlies those results. The tolerance
$\eta^{(k)} < 10^{-9}$ and the iteration cap of two thousand sweeps
adopted here are configuration-file parameters that may be adjusted at run
time.

Once the potentials $(f, g)$ satisfy the convergence
criterion~\eqref{eq:marginal_residual}, the optimal transport plan is
materialized on the exponential scale by direct evaluation of
\begin{equation}
\pi^\star_{ij} \;=\; \exp\!\left( \frac{f_i + g_j - C_{ij}}{\varepsilon} \right),
\qquad
i = 1, \ldots, n,\;
j = 1, \ldots, m,
\label{eq:plan_materialize}
\end{equation}
consistent with the Gibbs factorization~\eqref{eq:gibbs_factor} after the
change of variables~\eqref{eq:log_potentials}. Materialization is performed
only once, after the iterative refinement has terminated, and the
associated loss of log-domain stability is confined to individual matrix
entries whose value is below the smallest representable positive
double-precision number and which contribute negligibly to any downstream
statistic~\cite{schmitzer2019}. The transport cost is computed as the
Frobenius pairing
\begin{equation}
\langle C, \pi^\star \rangle_F \;=\; \sum_{i=1}^n \sum_{j=1}^m
C_{ij}\, \pi^\star_{ij},
\label{eq:transport_cost_final}
\end{equation}
evaluated on the materialized plan, matching the
objective~\eqref{eq:eot_discrete} evaluated at the optimum.

Reconstruction of the intermediate-time density $\rho_t$
in~\eqref{eq:rho_t} proceeds by drawing samples from the marginal law
rather than by evaluating $\rho_t$ on a spatial grid, which would be
infeasible for the domain-free point-cloud data type. The Brownian-bridge
formula~\eqref{eq:bridge_marginal} shows that a sample from $\rho_t$ may
be obtained by a two-step composition: first draw endpoint indices $(i,
j)$ from the discrete joint distribution induced by $\pi^\star$, and then
draw the intermediate position from the conditional Gaussian law with
mean $(1 - t)\, x_i + t\, y_j$ and covariance $\varepsilon\, t (1 - t)\,
I_d$. Conditioned on the source index $i$, the target index $j$ is drawn
from the discrete conditional distribution
\begin{equation}
\mathbb{P}( j \,|\, i ) \;=\; \frac{\pi^\star_{ij}}{\sum_{j'} \pi^\star_{ij'}}
\;=\; \frac{\pi^\star_{ij}}{a_i},
\label{eq:conditional_draw}
\end{equation}
where the second equality holds exactly at convergence and is the reason
that the marginal residual~\eqref{eq:marginal_residual} rather than the
plan itself controls sampling fidelity. The categorical draw is executed
by inverse cumulative distribution function sampling, that is, by drawing
a uniform random variable $U \sim \mathcal{U}(0, 1)$ and selecting the
smallest index $j^\star(i, U)$ for which
\begin{equation}
\sum_{j' = 1}^{j^\star} \frac{\pi^\star_{ij'}}{a_i} \;\geq\; U.
\label{eq:inverse_cdf}
\end{equation}
The intermediate-time sample is then
\begin{equation}
X_{t, i} \;=\; (1 - t)\, x_i \;+\; t\, y_{j^\star(i, U)} \;+\;
\sqrt{\varepsilon\, t (1 - t)}\; Z_i,
\qquad Z_i \sim \mathcal{N}(0, I_d),
\label{eq:bridge_sample}
\end{equation}
with the Gaussian noise drawn coordinatewise from the standard normal
distribution provided by \texttt{np.random.randn} and scaled by the bridge
standard deviation $\sqrt{\varepsilon\, t (1 - t)}$ derived
from~\eqref{eq:bridge_marginal}. A defensive branch reassigns the sample
to $x_i$ whenever the row sum $\sum_j \pi^\star_{ij}$ falls below
$10^{-300}$, an eventuality that in practice occurs only when the
convergence tolerance has been set orders of magnitude below the
achievable double-precision floor.

Full trajectories are constructed by repeating the sampling
procedure~\eqref{eq:inverse_cdf}--\eqref{eq:bridge_sample} over a uniform
time grid
\begin{equation}
t_\ell \;=\; \frac{\ell}{N_f - 1},
\qquad \ell = 0, 1, \ldots, N_f - 1,
\label{eq:time_grid}
\end{equation}
of $N_f$ frames on the unit interval. At $t = 0$ and $t = 1$ the boundary
constraints are enforced exactly by returning the source cloud $X$ and the
target cloud $Y$ respectively, bypassing the stochastic sampling procedure
and guaranteeing that the reconstructed trajectory reproduces the
prescribed endpoints to machine precision. At interior time levels the
sampling routine is called with a per-frame random seed constructed by
adding the frame index to a user-specified base seed, which yields
independent draws from $\rho_{t_\ell}$ across frames and produces a
visualization of marginal evolution rather than of Lagrangian particle
paths. While a pathwise construction, in which each particle is assigned
a fixed Brownian-bridge realization across all frames, would produce
smoother apparent trajectories, the marginal-sampling procedure adopted
here has the advantage that each frame is a statistically valid draw from
the correct intermediate density.

Beyond the primary transport plan and its bridge reconstruction, the
solver evaluates a small collection of diagnostic quantities at
convergence. The plan entropy
\begin{equation}
H(\pi^\star) \;=\; -\sum_{i=1}^n \sum_{j=1}^m \pi^\star_{ij}\,
\log \pi^\star_{ij}
\label{eq:plan_entropy}
\end{equation}
measures the effective dispersion of the coupling, evaluated with the
convention $0 \log 0 = 0$ and restricted in practice to entries above the
underflow floor of $10^{-300}$. The normalized effective support
$\exp(H(\pi^\star)) / (n m)$ translates the entropy into a scale-free
number in $(0, 1]$ that approaches unity for a diffuse near-independent
coupling and $1 / \max(n, m)$ for a permutation-like coupling that
concentrates all mass on a small subset of entries. The row and column
marginal residuals evaluated on the materialized
plan~\eqref{eq:plan_materialize},
\begin{equation}
r_{\mathrm{row}} \;=\; \max_i \left| \sum_{j=1}^m \pi^\star_{ij} - a_i \right|,
\qquad
r_{\mathrm{col}} \;=\; \max_j \left| \sum_{i=1}^n \pi^\star_{ij} - b_j \right|,
\label{eq:marginal_max_errors}
\end{equation}
together with the total plan mass $\sum_{ij} \pi^\star_{ij}$ and the
$\ell^1$ residual~\eqref{eq:marginal_residual} at the final iteration,
provide a consistency audit whose deviation from the theoretical values of
$0$, $0$, $1$, and $0$ respectively quantifies the numerical residual
accumulated during the iteration. All diagnostics, along with the
potentials, the plan, and the bridge trajectory, are written to a
self-describing NetCDF-4 file following the Climate and Forecast (CF)
conventions~\cite{rew1990,hoyer2017}, with \texttt{float32} storage adopted
for the bulk quantities to keep archival footprints manageable while
retaining the \texttt{float64} working precision at run time.

\subsection{Numerical Experiments}

Four demonstration cases exercise \texttt{anyakrakusuma} across a graded
sequence of source and target geometries. Each case fixes the point-cloud
size at $n = m = 1000$, the spatial dimension at $d = 2$, the marginal
weights at uniform values $a_i = 1/n$ and $b_j = 1/m$, the Sinkhorn
tolerance at $\eta^{(k)} < 10^{-9}$, the iteration cap at two thousand
sweeps, and the base random seed at the value $42$; source and target
point clouds are generated with random seeds $42$ and $1042$ respectively,
so that the two clouds are drawn from independent streams of a
reproducible pseudorandom sequence. The regularization parameter
$\varepsilon$ and the frame count $N_f$ are chosen case by case, and their
values are tabulated together with the source and target parametrizations
in the descriptions that follow. Although the demonstration parameters
adopt uniform marginals and a two-dimensional embedding, the underlying
routines described earlier accept nonuniform marginal weights and
arbitrary spatial dimension $d$; the built-in distribution generators and
the animation utilities of \texttt{anyakrakusuma}, however, are
specialized to the two-dimensional case, so that experiments beyond the
plane require user-supplied point clouds.

\paragraph{Case 1: Circle to circle.}
The source distribution is the uniform measure on the unit circle,
sampled by
\begin{equation}
x_i \;=\; \left( \cos \theta_i,\; \sin \theta_i \right),
\qquad
\theta_i \;=\; \frac{2 \pi (i - 1)}{n} + \varphi,
\qquad i = 1, \ldots, n,
\label{eq:case1_source}
\end{equation}
with a global phase $\varphi \sim \mathcal{U}(0, 2\pi/n)$ drawn once at
the outset to randomize the angular offset. The target distribution is
the uniform measure on the concentric circle of radius $2$, sampled by
the same construction with radius rescaled from $1$ to $2$ and an
independent phase drawn from the same distribution. No perturbation
noise is applied at either endpoint, so the two clouds lie exactly on
their respective circles. The regularization parameter is set to
$\varepsilon = 0.02$ and the trajectory is rendered on a grid of $N_f =
120$ frames. This case isolates radial rescaling in the absence of
angular reorganization and provides a baseline against which the more
complex geometries are compared.

\paragraph{Case 2: Spiral to Gaussian mixture.}
The source distribution is an Archimedean spiral with two full turns,
sampled by
\begin{equation}
x_i \;=\; r_i \left( \cos \theta_i,\; \sin \theta_i \right),
\qquad
\theta_i \;=\; 0.5 + \frac{4 \pi - 0.5}{n - 1}\, (i - 1),
\qquad
r_i \;=\; \frac{1.5\, \theta_i}{4 \pi},
\label{eq:case2_source}
\end{equation}
for $i = 1, \ldots, n$, with additive isotropic Gaussian perturbation of
standard deviation $\sigma_{\mathrm{src}} = 0.01$. The target distribution
is a mixture of four Gaussian components with centers arranged uniformly
on the circle of radius $1.5$,
\begin{equation}
c_k \;=\; 1.5\, \left( \cos \frac{2 \pi (k - 1)}{4},\;
\sin \frac{2 \pi (k - 1)}{4} \right),
\qquad k = 1, 2, 3, 4,
\label{eq:case2_target_centers}
\end{equation}
with each component contributing $n / 4$ points drawn from
$\mathcal{N}(c_k, 0.15^2 I_2)$. The regularization is set to $\varepsilon
= 0.05$ and the trajectory is rendered on $N_f = 120$ frames. This case
tests the solver on a topology-changing transition from a
one-dimensional connected support to a disconnected four-modal target.

\paragraph{Case 3: Two moons to rotated two moons.}
The source distribution is the two-moons configuration, an interleaved
pair of half-circles sampled by
\begin{equation}
x_i \;=\;
\begin{cases}
( \cos \alpha_i,\; \sin \alpha_i ) + \xi_i,
& i = 1, \ldots, \lfloor n / 2 \rfloor, \\[4pt]
( 1 - \cos \beta_i,\; \tfrac{1}{2} - \sin \beta_i ) + \xi_i,
& i = \lfloor n / 2 \rfloor + 1, \ldots, n,
\end{cases}
\label{eq:case3_source}
\end{equation}
where the angular parameters $\alpha_i \in [0, \pi]$ and $\beta_i \in [0,
\pi]$ are equispaced within their respective half of the sample and
$\xi_i \sim \mathcal{N}(0, 0.05^2 I_2)$ is the additive isotropic
perturbation applied independently to each point. The target
distribution is obtained by applying a rotation of $\pi/2$ radians about
the empirical centroid $\bar x = n^{-1} \sum_i x_i$ of a freshly drawn
two-moons cloud,
\begin{equation}
y_j \;=\; \bar x \;+\;
\begin{pmatrix}
\cos(\pi/2) & -\sin(\pi/2) \\
\sin(\pi/2) & \phantom{-}\cos(\pi/2)
\end{pmatrix}
( x'_j - \bar x ),
\qquad j = 1, \ldots, n,
\label{eq:case3_target}
\end{equation}
where $\{x'_j\}_{j=1}^n$ is the fresh two-moons cloud generated with the
target random seed. The regularization is set to $\varepsilon = 0.03$ and
the trajectory is rendered on $N_f = 120$ frames. This case tests the
coupling on a nontrivial angular reorganization between two clouds of
identical shape but distinct orientation.

\paragraph{Case 4: Lissajous curve to trefoil projection.}
The source distribution is the Lissajous curve with frequency ratio $3
{:} 2$ and phase shift $\pi/2$, scaled to unit amplitude $1.5$,
\begin{equation}
x_i \;=\; 1.5\, \left( \sin(3 t_i + \pi/2),\; \sin(2 t_i) \right),
\qquad
t_i \;=\; \frac{2 \pi (i - 1)}{n},
\qquad i = 1, \ldots, n,
\label{eq:case4_source}
\end{equation}
sampled uniformly in the parameter $t \in [0, 2\pi)$. The target
distribution is the two-dimensional projection of the trefoil knot,
\begin{equation}
y_j \;=\; \frac{1.5}{3}\, \left(
\sin s_j + 2 \sin(2 s_j),\;
\cos s_j - 2 \cos(2 s_j)
\right),
\qquad
s_j \;=\; \frac{2 \pi (j - 1)}{n},
\qquad j = 1, \ldots, n,
\label{eq:case4_target}
\end{equation}
sampled uniformly in the parameter $s \in [0, 2\pi)$. No perturbation
noise is applied to either curve. The regularization is set to
$\varepsilon = 0.04$ and the trajectory is rendered on the denser grid of
$N_f = 150$ frames, reflecting the greater topological complexity of the
intermediate density in the transition between the two self-intersecting
curves.

\subsection{Data Analyses}

The Sinkhorn iteration history, the optimal coupling, and the bridge
trajectory stored in each NetCDF file are subjected to four independent
diagnostic analyses that quantify, respectively, the numerical convergence
of the dual iteration, the geometric content of the discrete coupling, the
spatial evolution of the intermediate density, and the informational and
structural evolution of the point cloud along the bridge. All four
analyses are implemented in Python and rely on \texttt{NumPy} for array
manipulation~\cite{harris2020}, \texttt{SciPy} for the kernel density
estimator, nearest-neighbor queries, and special
functions~\cite{virtanen2020}, and \texttt{Matplotlib} for the
publication figures~\cite{hunter2007}, with the NetCDF interface exposed
through \texttt{netCDF4}~\cite{rew1990}. The four diagnostic categories
are treated in turn below.

The convergence history of the log-domain Sinkhorn iteration is recorded
by the solver every ten sweeps, so that the vector $\{\eta^{(k_\ell)}\}$ of
stored marginal residuals is associated with iteration indices
\begin{equation}
k_\ell \;=\; 10\, \ell,
\qquad \ell = 0, 1, \ldots, L - 1,
\label{eq:iter_axis}
\end{equation}
where $L$ is the number of stored samples and $\eta^{(k)}$ is the row
marginal violation defined in~\eqref{eq:marginal_residual}. Under the
contraction properties of the Sinkhorn recursion in the Hilbert
projective metric~\cite{sinkhorn1964,knight2008}, the residual is expected
to decay geometrically, so that $\log_{10} \eta^{(k)}$ is approximately
linear in $k$. A least-squares fit
\begin{equation}
\log_{10} \eta^{(k_\ell)} \;\approx\; \beta_0 + \beta_1\, k_\ell,
\qquad \ell = 0, 1, \ldots, L - 1,
\label{eq:log_linear_fit}
\end{equation}
is carried out over the positive-residual subset, and the per-iteration
contraction factor is reported as $10^{\beta_1}$. As an independent check
of the geometric-decay assumption, a per-step contraction ratio
\begin{equation}
r^{(k_\ell)} \;=\; \left( \frac{\eta^{(k_{\ell+1})}}{\eta^{(k_\ell)}} \right)^{1/10},
\qquad \ell = 0, 1, \ldots, L - 2,
\label{eq:contraction_ratio}
\end{equation}
is computed and plotted against $k_\ell$, with the exponent $1/10$
converting the between-record ratio to a per-iteration quantity. A flat
sequence $r^{(k_\ell)}$ approximately equal to $10^{\beta_1}$ indicates
that the geometric-decay assumption is well satisfied, while systematic
drift signals a departure from strict linearity in the log-domain. The
fit~\eqref{eq:log_linear_fit} is defined only when at least two
positive-residual samples are recorded, so scenarios that converge below
tolerance at the first evaluation contribute one summary line to the
report but no fitted rate.

The optimal coupling $\pi^\star$ materialized through~\eqref{eq:plan_materialize}
is analyzed through the discrete conditional distribution and the
associated barycentric projection. For each source index $i$, the row
conditional
\begin{equation}
P(j \,|\, i) \;=\; \frac{\pi^\star_{ij}}{\sum_{j' = 1}^m \pi^\star_{ij'}},
\qquad j = 1, \ldots, m,
\label{eq:row_conditional}
\end{equation}
which coincides with the sampling distribution~\eqref{eq:conditional_draw}
used in the bridge reconstruction, gives the entropic-transport analogue
of the target of a deterministic Monge map. The barycentric image of
$x_i$ under the coupling is the conditional-expectation map
\begin{equation}
T(x_i) \;=\; \mathbb{E}[Y \,|\, X = x_i]
\;=\; \sum_{j = 1}^m P(j \,|\, i)\, y_j,
\qquad i = 1, \ldots, n,
\label{eq:barycentric_map}
\end{equation}
which in the deterministic limit $\varepsilon \to 0^+$ approaches the
Brenier map and in the diffuse limit $\varepsilon \to \infty$ collapses to
the marginal mean of the target~\cite{brenier1991,peyrecuturi2019}. The
per-source dispersion of the coupling is quantified through the row
Shannon entropy
\begin{equation}
H_i \;=\; -\sum_{j = 1}^m P(j \,|\, i)\, \log P(j \,|\, i),
\qquad i = 1, \ldots, n,
\label{eq:row_entropy}
\end{equation}
with the convention $0 \log 0 = 0$ enforced by restricting the sum to the
positive-mass entries. The corresponding perplexity $\exp(H_i)$ is the
effective number of target points that carry nonvanishing mass from
$x_i$, and its arithmetic mean $\langle \exp(H_i) \rangle_i$ is reported
as a single case-level summary of the coupling's diffuseness. A
complementary summary is the mean peak conditional probability $\langle
\max_j P(j \,|\, i) \rangle_i$, which approaches unity for permutation-like
plans and $1/m$ for the diffuse product coupling. The displacement
magnitude $\|T(x_i) - x_i\|$ furnishes a scalar per-source measure of the
transport work, reported as its mean, median, and maximum across the
source cloud. When the stored plan is subsampled to a $p \times p$
retained block by the solver's NetCDF writer, the source and target
clouds are aligned to the retained rows and columns through the same
strided index selection $\iota_\ell = \lfloor \ell (n - 1) / (p - 1)
\rfloor$ used at storage time, and the retained rows are renormalized
before the conditional~\eqref{eq:row_conditional} is formed. In this
subsampled regime the reported quantities are computed over a strided
subset rather than over the full plan and are labeled accordingly in the
diagnostic report.

The intermediate marginals $\rho_t$ reconstructed through the bridge
sampler are subjected to a Gaussian kernel density estimate (KDE) at five
representative interpolation times $\mathcal{T} = \{0,\, 0.25,\, 0.5,\,
0.75,\, 1\}$. For a snapshot time $t^\star \in \mathcal{T}$, the sample
cloud is either the stored frame at $t^\star$ when the time grid contains
$t^\star$ exactly, or the linear temporal interpolation between the two
straddling frames when it does not. On the resulting cloud, the density
estimate at grid point $z$ is
\begin{equation}
\hat \rho_t(z) \;=\; \frac{1}{n\, h^d}\, \sum_{i = 1}^n
K\!\left( \frac{z - X_{t, i}}{h} \right),
\label{eq:kde}
\end{equation}
with an isotropic Gaussian kernel $K$ and bandwidth $h$ selected by
Scott's rule of thumb, $h = n^{-1/(d + 4)}$~\cite{scott1979,silverman1986},
as implemented in the \texttt{scipy.stats.gaussian\_kde}
routine~\cite{virtanen2020}. The density is evaluated on a $140 \times 140$
grid spanning the joint bounding box of the source, target, and all
intermediate clouds, and is normalized to its per-panel maximum so that
the cross-time and cross-case comparison is carried out on the
scale-free field $\rho^\dagger_t(z) = \hat \rho_t(z) / \max_z \hat
\rho_t(z)$. Three summary quantities are computed on $\rho^\dagger_t$.
The high-density region count
\begin{equation}
N_{\mathrm{reg}}(t) \;=\; \#\, \big\{ \text{connected components of }
\{ z : \rho^\dagger_t(z) \geq 0.5 \} \big\},
\label{eq:region_count}
\end{equation}
formed under four-connectivity on the pixel lattice, provides a
ridge-robust replacement for the raw mode count: a ring-shaped support
counts as a single region, while spatially separated concentrations count
individually. The effective support area
\begin{equation}
A_{\mathrm{eff}}(t) \;=\; \exp\!\left( - \sum_{c} p_c(t)\, \log p_c(t) \right)
\Delta x\, \Delta y,
\qquad p_c(t) \;=\; \frac{\rho^\dagger_t(z_c)}{\sum_{c'} \rho^\dagger_t(z_{c'})},
\label{eq:effective_area}
\end{equation}
in which the sum runs over grid cells $c$ with center $z_c$ and area
$\Delta x\, \Delta y$, gives a participation-based measure of the area
occupied by the density that does not degenerate for supports concentrated
near a low-dimensional manifold, in contrast to the point-mass summary
that would be given by the argmax of $\rho^\dagger_t$. The support
fraction $| \{ z : \rho^\dagger_t(z) \geq 0.03 \} | / (140^2)$ complements
the effective area with a simple thresholded-coverage statistic that
tracks with the visual footprint of each panel.

The bridge point cloud is further characterized at every stored frame
through the Kozachenko--Leonenko $k$-nearest-neighbor estimator of the
differential entropy of $\rho_t$~\cite{kozachenko1987,kraskov2004},
\begin{equation}
\hat H(\rho_t) \;=\; -\psi(k) + \psi(n) + \log c_d
\;+\; \frac{d}{n} \sum_{i = 1}^n \log r_i(t),
\label{eq:kl_entropy}
\end{equation}
with $k = 5$, digamma function $\psi$, unit-ball volume
\begin{equation}
c_d \;=\; \frac{\pi^{d/2}}{\Gamma(d/2 + 1)},
\label{eq:unit_ball}
\end{equation}
and $r_i(t)$ the Euclidean distance from $X_{t, i}$ to its $k$-th nearest
neighbor in the frame-$t$ cloud, computed through the
\texttt{scipy.spatial.cKDTree} query with a floor of $10^{-12}$ to guard
against exact-coincidence pairs. The estimator~\eqref{eq:kl_entropy} is
consistent for absolutely continuous densities in $\mathbb{R}^d$ and
exhibits $\sqrt{n}$-convergence in the limit of large sample size; its
finite-sample behavior at the endpoint clouds, where the mass concentrates
near a lower-dimensional curve, is understood as a comparative diagnostic
rather than as an estimate of a well-defined two-dimensional differential
entropy. As a reference against which the measured value at the midpoint
of the bridge may be compared, the differential entropy of the pure
Brownian-bridge noise term of~\eqref{eq:bridge_marginal}, whose covariance
is $\varepsilon\, t (1 - t)\, I_d$, is
\begin{equation}
H_{\mathrm{diff}}(t) \;=\; \frac{d}{2}\, \log\!\left(
2 \pi e\, \varepsilon\, t (1 - t) \right),
\label{eq:diffusive_reference}
\end{equation}
which at $t = 1/2$ reduces to $(d/2) \log(\pi e\, \varepsilon / 2)$ and
serves as a lower bound on the entropy that a bridge marginal must carry
whenever the endpoint contribution is negligible.

The second-moment geometry of $\rho_t$ is summarized through the sample
covariance matrix
\begin{equation}
\hat \Sigma(t) \;=\; \frac{1}{n - 1} \sum_{i = 1}^n
\left( X_{t, i} - \bar X_t \right) \left( X_{t, i} - \bar X_t \right)^\top,
\qquad
\bar X_t \;=\; \frac{1}{n} \sum_{i = 1}^n X_{t, i},
\label{eq:sample_covariance}
\end{equation}
with eigenvalues $\lambda_{\min}(t) \leq \lambda_{\max}(t)$ computed
through \texttt{numpy.linalg.eigvalsh}. Three ellipse descriptors follow:
the root-mean-square (RMS) dispersion
\begin{equation}
\mathrm{rms}(t) \;=\; \sqrt{\mathrm{tr}\, \hat \Sigma(t)}
\;=\; \sqrt{\lambda_{\min}(t) + \lambda_{\max}(t)},
\label{eq:rms_dispersion}
\end{equation}
the eccentricity
\begin{equation}
e(t) \;=\; \sqrt{ 1 - \frac{\lambda_{\min}(t)}{\lambda_{\max}(t)} },
\label{eq:eccentricity}
\end{equation}
and the doubled principal-axis angle
\begin{equation}
\Theta(t) \;=\; \operatorname{atan2}\!\left(
2\, \hat \Sigma_{12}(t),\;
\hat \Sigma_{11}(t) - \hat \Sigma_{22}(t)
\right),
\label{eq:doubled_angle}
\end{equation}
whose halving gives the principal-axis direction modulo $\pi$. The
doubled representation removes the axis-wrap ambiguity that would
corrupt a direct estimate of the axis angle, and permits temporal
unwrapping to be performed by standard phase-unwrap on
$\Theta(t)$~\cite{mardia2000}. Because the principal-axis direction is
resolved only for anisotropic clouds, the orientation angle is
reported only at frames with eccentricity above a fixed threshold, $e(t)
\geq 0.25$; below this threshold the two eigenvalues are too close for
the principal axis to be meaningfully distinguished from a random
rotation of an isotropic disk. The unwrap is applied within each
contiguous run of resolved frames but never across a masked run, since
the axis direction on the two sides of an isotropic gap can be
near-antipodal and the connecting branch is not determined by the data.

Uncertainty in the frame-wise estimates of the differential
entropy~\eqref{eq:kl_entropy}, the RMS dispersion~\eqref{eq:rms_dispersion},
the eccentricity~\eqref{eq:eccentricity}, and the doubled
angle~\eqref{eq:doubled_angle} is quantified through subsampling without
replacement, chosen in preference to the classical bootstrap because
resampling with replacement generates coincident particles that corrupt
the nearest-neighbor entropy through $\log 0$ singularities in the
$r_i$~\cite{politis1994,efron1993}. For each frame, $B = 80$ subsamples
of size $m = \lfloor 0.8\, n \rfloor$ are drawn from the frame cloud
without replacement, each of the four descriptors is recomputed on the
subsample, and the subsample standard deviation is rescaled to the
full-sample standard deviation through the $\sqrt{n}$-convergence
identity
\begin{equation}
\hat \sigma_n \;=\; \sqrt{\frac{m}{n}}\; \hat \sigma_m,
\label{eq:subsampling_rescale}
\end{equation}
which is the standard subsampling correction for statistics whose
asymptotic distribution is centered at a fixed limit and scales as
$n^{-1/2}$~\cite{politis1994}. A ninety-five-percent confidence band is
then formed as $\hat \theta_n \pm 1.96\, \hat \sigma_n$ under a Gaussian
approximation for the linear descriptors. For the doubled angle
$\Theta$, whose ambient space is the unit circle, the subsample spread
is quantified through the mean resultant length
\begin{equation}
R \;=\; \left| \frac{1}{B} \sum_{b = 1}^B \exp\!\left( i\, \Theta^{(b)} \right) \right|,
\label{eq:mean_resultant}
\end{equation}
from which the circular standard deviation follows through the standard
identity $\sigma_{\mathrm{circ}} = \sqrt{-2 \log
R}$~\cite{mardia2000}, undefined in the vanishing-$R$ limit and guarded
in the implementation by a lower cutoff $R > 10^{-12}$. The
corresponding half-width on the axis angle, obtained after halving the
doubled representation and rescaling through~\eqref{eq:subsampling_rescale},
is reported in degrees.

The per-case diagnostic reports collect these quantities into a fixed
plain-text layout that lists, for each of the four cases, the
convergence flag and iteration count, the log-linear contraction factor
and its per-step check, the barycentric-map summary
statistics, the KDE snapshot table with region count, effective area,
support fraction, and centroid at each of the five snapshot times, and
the frame-wise differential entropy at $t \in \{0, 1/2, 1\}$ together
with the net production $\hat H(\rho_1) - \hat H(\rho_0)$, the peak
entropy and its time of occurrence, the RMS dispersion at the three
canonical times together with its peak, and the axis reorientation
$(\Theta(1) - \Theta(0) + 90^\circ) \bmod 180^\circ - 90^\circ$ folded to
the interval $(-90^\circ,\, 90^\circ]$ to remove the mod-$\pi$ ambiguity
of principal-axis directions. Each descriptor is accompanied by the mean
across frames of the ninety-five-percent half-width from the subsampling
calculation, which quantifies the average size of the uncertainty band
in the corresponding figure panel and provides a scalar summary of the
statistical resolution at which each geometric feature has been
recovered.

\section{Results}

The four demonstration cases were executed under the parameter settings
specified above, each as an independent process that logged its parameters,
solver diagnostics, and a timing breakdown alongside the NetCDF archive. The
resulting archives were passed through the four diagnostic analyses without
further intervention. All values reported below are taken directly from the
run logs and the diagnostic output. Quantities are dimensionless unless a
unit is stated.

Every case satisfied the marginal-residual tolerance $\eta^{(k)} < 10^{-9}$
within the two-thousand-sweep cap, and every run reported a marginal fidelity
of $1.00000000$ and terminated without warnings or errors. The sweep counts
at termination were $1$, $721$, $581$, and $651$ for cases~1 through~4
respectively, corresponding to $1$, $73$, $59$, and $66$ recorded residual
samples at the ten-sweep recording stride. Case~1 met the tolerance at the
first residual evaluation, at which point the residual stood at $\eta =
3.771506 \times 10^{-15}$, approximately seventeen times the double-precision
unit roundoff $2^{-52} \approx 2.220 \times 10^{-16}$. The log-linear fit of
equation~\eqref{eq:log_linear_fit} is undefined for a single recorded
sample, so no contraction factor is reported for case~1 and that case is
absent from the convergence panels. The remaining three cases entered the
iteration with residuals of order unity, $\eta^{(0)} = 1.127457$, $0.799201$,
and $0.535817$ for cases~2, 3, and~4, and terminated at $8.116460 \times
10^{-10}$, $9.982731 \times 10^{-10}$, and $8.503125 \times 10^{-10}$. The
fitted slopes were $-1.071160 \times 10^{-2}$, $-1.511440 \times 10^{-2}$,
and $-1.171978 \times 10^{-2}$ decades per sweep, giving per-iteration
contraction factors $10^{\beta_1}$ of $0.975637$, $0.965796$, and
$0.973375$. The ordering of the contraction factors across the three fitted
cases does not follow the ordering of the regularization parameter, the
fastest contraction being recorded at $\varepsilon = 0.03$ and the slowest at
$\varepsilon = 0.05$. Table~\ref{tab:convergence} collects the convergence
summary.

\begin{figure}[H]
\centering
\includegraphics[width=\linewidth]{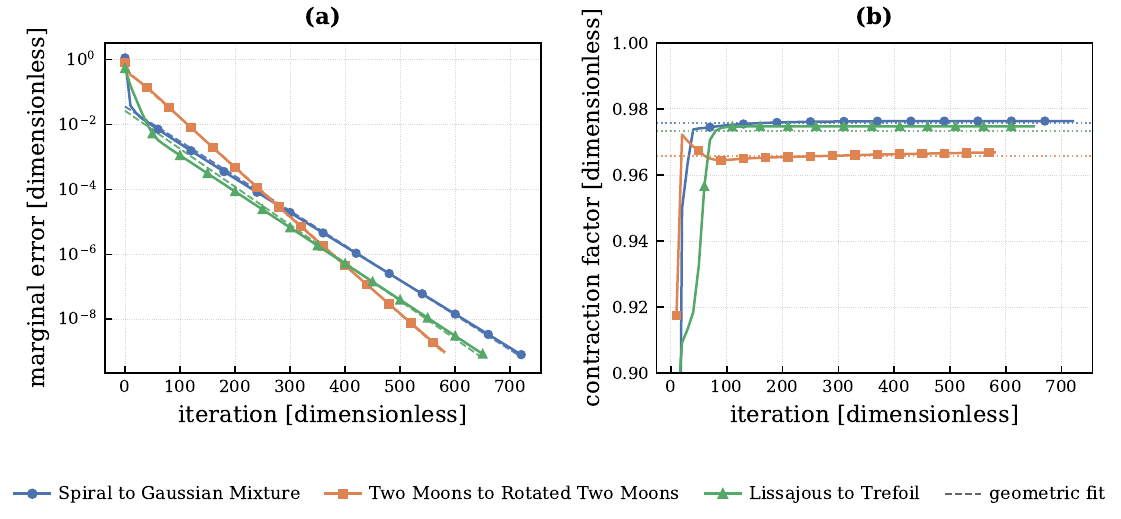}
\caption{Convergence of the log-domain Sinkhorn iteration for the three cases
that produced a fittable residual history. (a) Marginal constraint violation
$\eta^{(k)} = \| \pi^{(k)} \mathbf{1}_m - a \|_1$ against sweep index, with
the fitted geometric decays of equation~\eqref{eq:log_linear_fit} overlaid as
dashed lines. (b) Per-iteration contraction ratio $r^{(k_\ell)}$ of
equation~\eqref{eq:contraction_ratio} against sweep index, with the fitted
factor $10^{\beta_1}$ of each case shown as a dotted horizontal line. Case~1
satisfied the convergence tolerance at the first residual evaluation and
contributes a single recorded sample, from which no rate can be fitted; it is
therefore absent from both panels.}
\label{fig:convergence}
\end{figure}

Figure~\ref{fig:convergence}(a) shows the residual histories on a logarithmic
ordinate together with the fitted geometric decays. The histories are linear
over approximately eight decades in all three fitted cases, and the fitted
lines are visually indistinguishable from the data over the greater part of
that range. Figure~\ref{fig:convergence}(b) shows the per-step contraction
ratio of equation~\eqref{eq:contraction_ratio}. In each case the ratio rises
through a transient occupying the first several tens of sweeps and settles
onto the corresponding fitted factor within approximately one hundred sweeps,
remaining flat to within the resolution of the panel thereafter.

\begin{table}[H]
\centering
\caption{Convergence summary of the log-domain Sinkhorn iteration. The
residual $\eta^{(k)}$ is the $\ell^1$ row-marginal violation of
equation~\eqref{eq:marginal_residual}, recorded every ten sweeps. The
contraction factor is $10^{\beta_1}$ from the fit of
equation~\eqref{eq:log_linear_fit}.}
\label{tab:convergence}
\begin{tabular}{lccccc}
\toprule
Case & $\varepsilon$ & Sweeps & $\eta^{(0)}$ & $\eta^{\mathrm{final}}$ & $10^{\beta_1}$ \\
\midrule
1 & 0.02 & 1   & $3.7715 \times 10^{-15}$ & $3.7715 \times 10^{-15}$ & --- \\
2 & 0.05 & 721 & $1.1275 \times 10^{0}$   & $8.1165 \times 10^{-10}$ & 0.97564 \\
3 & 0.03 & 581 & $7.9920 \times 10^{-1}$  & $9.9827 \times 10^{-10}$ & 0.96580 \\
4 & 0.04 & 651 & $5.3582 \times 10^{-1}$  & $8.5031 \times 10^{-10}$ & 0.97338 \\
\bottomrule
\end{tabular}
\end{table}

Turning to the converged coupling, the transport cost $\langle C, \pi^\star
\rangle$ of equation~\eqref{eq:transport_cost_final} was $1.010013$,
$0.808137$, $0.866267$, and $0.313295$ for cases~1 through~4. The joint plan
entropy $H(\pi^\star)$ of equation~\eqref{eq:plan_entropy} was $10.7487$,
$11.6086$, $11.2869$, and $10.9313$ nats, and the effective sparsity
$\exp(H(\pi^\star))/(nm)$, the joint perplexity expressed as a fraction of the
$nm$ available cells, was $0.046568$, $0.110043$, $0.079772$, and $0.055899$.
The largest transport cost accompanies the largest interpolation distance and
the smallest coupling entropy accompanies the smallest regularization
parameter, with case~1 the sole exception in which the entropy is not the
largest despite the smallest cost.

The point-cloud size $n = m = 1000$ exceeds the $500 \times 500$ retention
limit of the NetCDF writer, so the archived plan is a strided block of the
full coupling in all four cases, and the conditional quantities that follow
are formed on that retained block after row renormalization. The mean row
entropy $\langle H_i \rangle_i$ of equation~\eqref{eq:row_entropy} was
$3.147765$, $4.008085$, $3.688627$, and $3.330434$ nats for cases~1 through
4, with across-row standard deviations of $4.15 \times 10^{-4}$, $0.341770$,
$0.193910$, and $0.249708$ nats. The across-row dispersion in case~1 is
therefore smaller than in the other three cases by between two and three
orders of magnitude. The corresponding mean perplexities $\langle \exp(H_i)
\rangle_i$ were $23.2840$, $58.0584$, $40.7015$, and $28.9141$ retained
target points, and the mean peak conditional probabilities $\langle \max_j
P(j\,|\,i) \rangle_i$ were $0.070853$, $0.047862$, $0.054963$, and
$0.065483$.

The block-normalized plan entropies recomputed on the retained block were
$9.362373$, $10.225071$, $9.898303$, and $9.545067$ nats. The difference
between the stored full-plan entropy and the block value takes the values
$1.386295$, $1.383553$, $1.388623$, and $1.386243$ nats and agrees with
$\log 4 = 1.386294$ to within $2.8 \times 10^{-3}$ nats in every case. The
barycentric displacement magnitudes $\| T(x_i) - x_i \|$ of
equation~\eqref{eq:barycentric_map} had means of $0.994994$, $0.819043$,
$0.822200$, and $0.467656$, medians of $0.994994$, $0.828984$, $0.803999$,
and $0.476303$, and maxima of $0.995037$, $1.232309$, $1.540545$, and
$0.989937$. In case~1 the mean, median, and maximum agree to within $4.3
\times 10^{-5}$, while in the remaining cases the maximum exceeds the mean by
between $0.41$ and $0.72$. Table~\ref{tab:coupling} collects the coupling
summary.

\begin{figure}[H]
\centering
\includegraphics[width=0.86\linewidth]{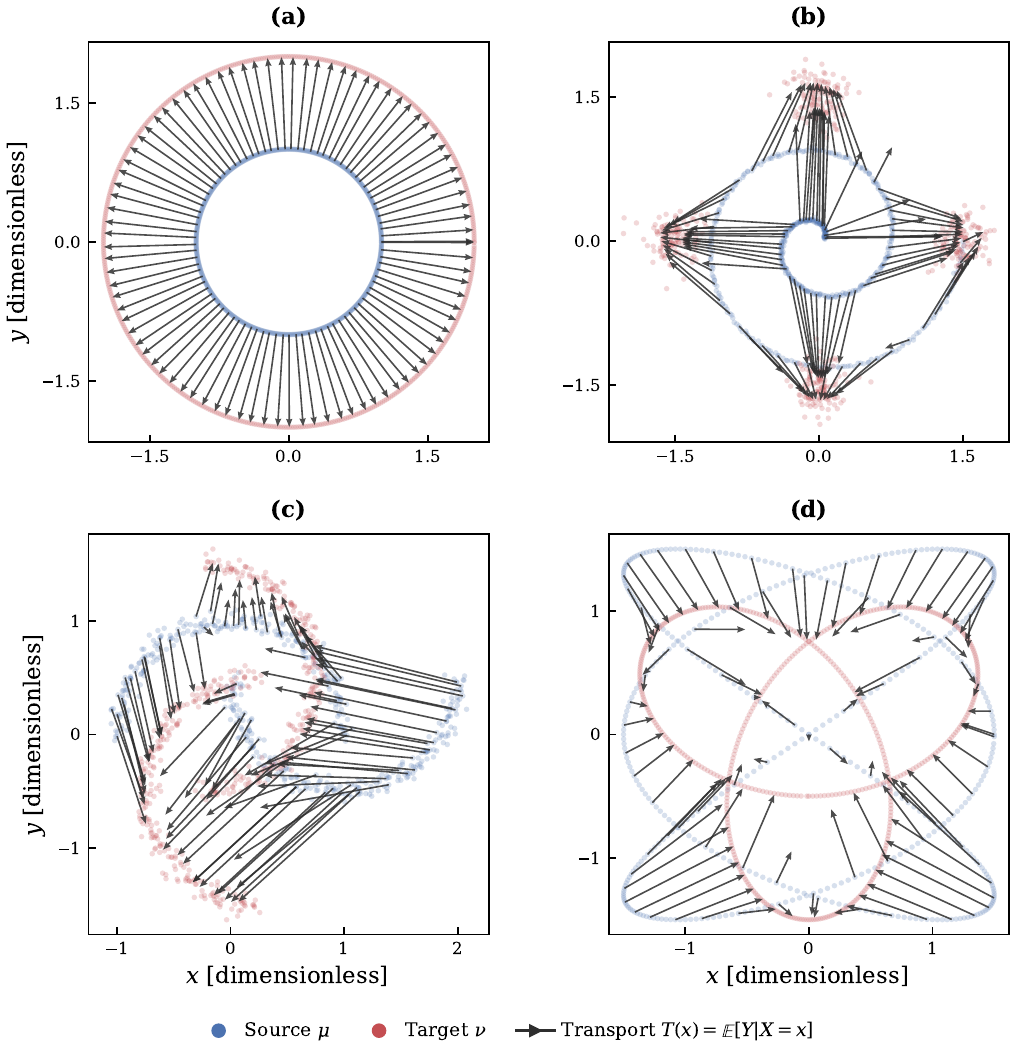}
\caption{Barycentric projection $T(x) = \mathbb{E}[Y \,|\, X = x]$ of the
entropic coupling, defined in equation~\eqref{eq:barycentric_map} and shown
as displacement arrows from source points to their conditional-mean images.
(a) Case~1, circle to circle. (b) Case~2, spiral to Gaussian mixture. (c)
Case~3, two moons to rotated two moons. (d) Case~4, Lissajous curve to
trefoil projection. Source and target clouds are drawn as faint markers.
Arrows are shown for a strided subset of ninety source points per panel for
legibility. Because $n = 1000$ exceeds the $500 \times 500$ storage retention
limit, the map is evaluated on the retained strided block of the coupling
after row renormalization.}
\label{fig:coupling_map}
\end{figure}

Figure~\ref{fig:coupling_map} shows the barycentric map of each case as
displacement arrows from a strided subset of ninety source points to their
conditional-mean images, overlaid on the source and target clouds. The arrows
in figure~\ref{fig:coupling_map}(a) are directed radially outward and are of
visually uniform length. Those in figure~\ref{fig:coupling_map}(b) converge
onto the four mixture centers, the arrow bundles originating along the turns
of the spiral and terminating in tight clusters.
Figure~\ref{fig:coupling_map}(c) shows arrows of markedly heterogeneous
length and direction. Figure~\ref{fig:coupling_map}(d) shows arrows directed
predominantly inward from the outer excursions of the Lissajous curve toward
the trefoil lobes.

\begin{table}[H]
\centering
\caption{Transport, conditional, and barycentric summary of the optimal
coupling. The transport cost $\langle C, \pi^\star \rangle$ and effective
sparsity $\exp(H(\pi^\star))/(nm)$ are evaluated on the full plan; the
conditional quantities are evaluated on the retained $500 \times 500$ block
after row renormalization, and the perplexity is in units of retained target
points.}
\label{tab:coupling}
\begin{tabular}{lcccccc}
\toprule
Case & $\varepsilon$ & $\langle C, \pi^\star \rangle$ & Eff. sparsity & $\langle H_i \rangle_i$ [nats] & $\langle \exp(H_i) \rangle_i$ & $\langle \| T(x) - x \| \rangle$ \\
\midrule
1 & 0.02 & 1.010013 & 0.046568 & 3.147765 & 23.2840 & 0.994994 \\
2 & 0.05 & 0.808137 & 0.110043 & 4.008085 & 58.0584 & 0.819043 \\
3 & 0.03 & 0.866267 & 0.079772 & 3.688627 & 40.7015 & 0.822200 \\
4 & 0.04 & 0.313295 & 0.055899 & 3.330434 & 28.9141 & 0.467656 \\
\bottomrule
\end{tabular}
\end{table}

The bridge marginals reconstructed from these couplings were examined next
through their KDEs. The Scott bandwidth factor returned
by the estimator was $0.3162$ in all four cases, consistent with
$n^{-1/(d+4)}$ at $n = 1000$ and $d = 2$. Of the five snapshot times
$\mathcal{T} = \{0,\, 0.25,\, 0.5,\, 0.75,\, 1\}$, the two endpoints coincide
with stored frames in every case, whereas the three interior times fall
between stored frames on both the $120$-frame and the $150$-frame grid and
were therefore obtained by linear temporal interpolation between the
straddling frames. The high-density region count $N_{\mathrm{reg}}(t)$ of
equation~\eqref{eq:region_count} held at unity across all five snapshots in
case~1 and at two across all five snapshots in case~3. In case~2 the count
followed the sequence $1,\, 3,\, 4,\, 4,\, 4$, and in case~4 the sequence
$2,\, 1,\, 1,\, 1,\, 1$. The effective support area $A_{\mathrm{eff}}(t)$ of
equation~\eqref{eq:effective_area} increased monotonically in case~1 from
$5.65001$ to $14.57646$, rose and then fell in cases~2 and~3 with maxima of
$8.39379$ at $t = 0.75$ and $6.18053$ at $t = 0.5$ respectively, and
decreased monotonically in case~4 from $11.28943$ to $8.40737$. The support
fraction reached the saturation value $1.00000$ at $t = 0$ and $t = 0.25$ in
case~4, and remained at or below $0.91041$ at every snapshot in the other
three cases.

The cloud centroid coincided with the origin to five decimal places at both
endpoints in cases~1 and~4, and departed from the origin by at most $1.98
\times 10^{-3}$ at the interior snapshots of case~1. In case~2 the centroid
moved from $(-0.00005,\, -0.12400)$ at $t = 0$ to $(-0.00787,\, 0.00398)$ at
$t = 1$. In case~3 the centroid moved from $(0.50166,\, 0.25285)$ at $t = 0$
to $(0.00000,\, 0.00000)$ at $t = 1$, a net displacement of magnitude
$0.56178$, with the intermediate values decreasing approximately linearly in
$t$. Table~\ref{tab:density} collects the snapshot summary.

\begin{figure}[H]
\centering
\includegraphics[width=\linewidth]{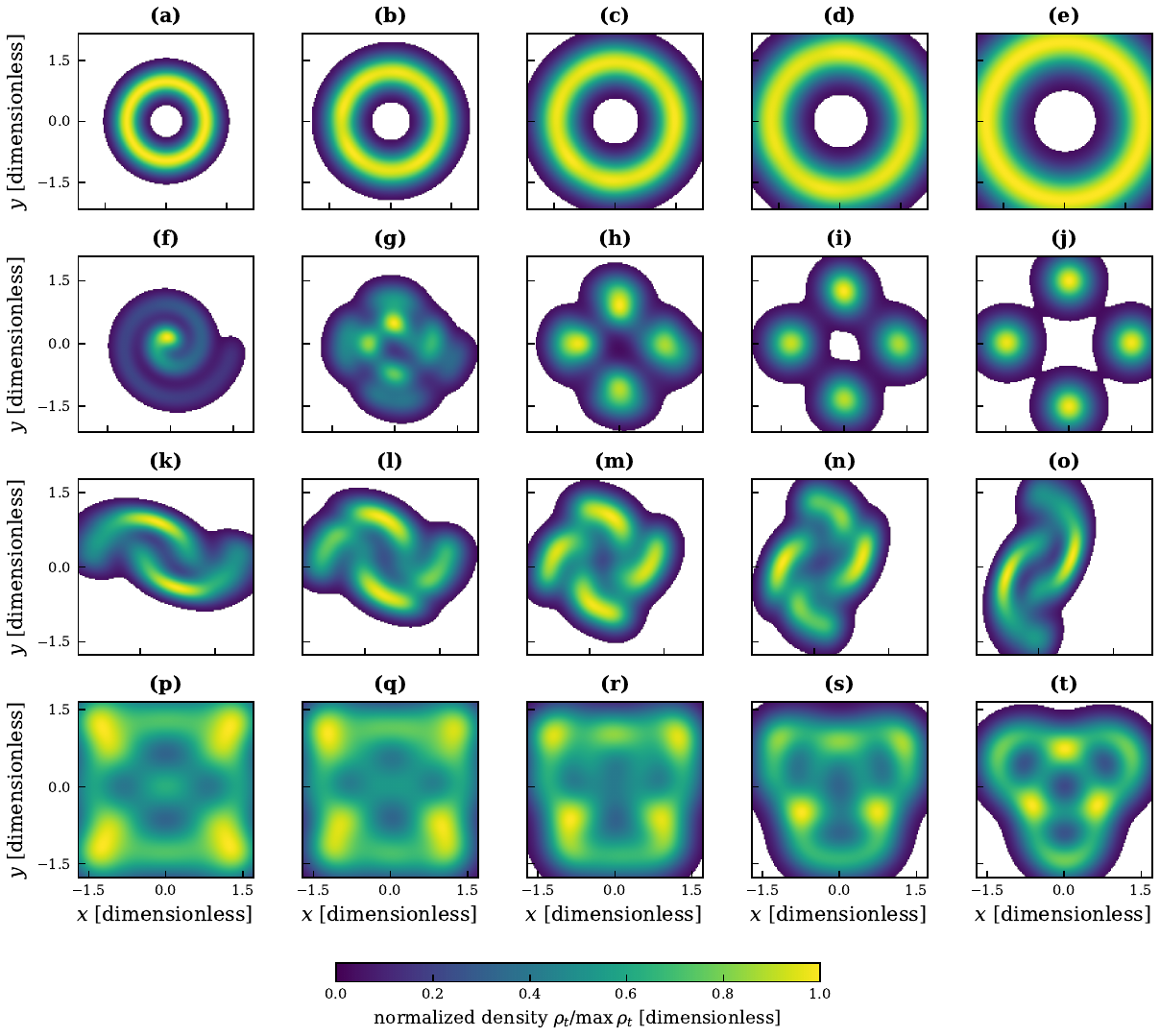}
\caption{Gaussian kernel density estimates of the bridge marginal $\rho_t$,
normalized to the per-panel maximum and evaluated on a $140 \times 140$ grid.
Rows correspond to cases~1 through~4 from top to bottom, columns to the
snapshot times $t = 0$, $0.25$, $0.5$, $0.75$, and $1$ from left to right.
Panels (a)--(e) show case~1, (f)--(j) case~2, (k)--(o) case~3, and (p)--(t)
case~4. Normalized density below $0.03$ is rendered white. The spatial extent
of each row is the joint bounding box of the source, target, and intermediate
clouds of that case and therefore differs between rows.}
\label{fig:density_evolution}
\end{figure}

Figure~\ref{fig:density_evolution} shows the normalized density
$\rho^\dagger_t$ as a four-by-five montage, one row per case and one column
per snapshot time. The first row shows an annulus of increasing radius and
decreasing relative thickness. The second row shows the spiral fragmenting
into four separated concentrations. The third row shows two crescent-shaped
ridges whose long axes differ in orientation between the first and the last
column. The fourth row shows four bright concentrations at $t = 0$ evolving
toward a three-lobed pattern at $t = 1$. The spatial extent of each row is
the joint bounding box of the clouds of that case, so panel areas are not
comparable between rows.

\begin{table}[H]
\centering
\caption{Kernel density estimate summary at the five snapshot times.
$N_{\mathrm{reg}}$ is the high-density region count of
equation~\eqref{eq:region_count}, $A_{\mathrm{eff}}$ the effective support
area of equation~\eqref{eq:effective_area}, and the support fraction the
proportion of the per-case panel at which the normalized density is at least
$0.03$. Effective areas and support fractions are referred to the per-case
bounding box and are not comparable across cases.}
\label{tab:density}
\begin{tabular}{llcccc}
\toprule
Case & $t$ & $N_{\mathrm{reg}}$ & $A_{\mathrm{eff}}$ & Support fraction & Centroid $(\bar x, \bar y)$ \\
\midrule
1 & 0.00 & 1 & 5.65001  & 0.38735 & $(0.00000,\, 0.00000)$ \\
  & 0.25 & 1 & 8.92757  & 0.61148 & $(-0.00196,\, 0.00031)$ \\
  & 0.50 & 1 & 12.25540 & 0.82862 & $(0.00017,\, -0.00003)$ \\
  & 0.75 & 1 & 14.25660 & 0.91000 & $(0.00170,\, 0.00020)$ \\
  & 1.00 & 1 & 14.57646 & 0.91041 & $(0.00000,\, 0.00000)$ \\
\midrule
2 & 0.00 & 1 & 6.05907  & 0.40842 & $(-0.00005,\, -0.12400)$ \\
  & 0.25 & 3 & 7.86339  & 0.53005 & $(-0.00586,\, -0.09478)$ \\
  & 0.50 & 4 & 8.38575  & 0.61194 & $(-0.01107,\, -0.06721)$ \\
  & 0.75 & 4 & 8.39379  & 0.63842 & $(-0.01332,\, -0.02325)$ \\
  & 1.00 & 4 & 7.92023  & 0.60842 & $(-0.00787,\, 0.00398)$ \\
\midrule
3 & 0.00 & 2 & 5.01334  & 0.48020 & $(0.50166,\, 0.25285)$ \\
  & 0.25 & 2 & 5.94372  & 0.57036 & $(0.37584,\, 0.19356)$ \\
  & 0.50 & 2 & 6.18053  & 0.59740 & $(0.25089,\, 0.12248)$ \\
  & 0.75 & 2 & 5.90077  & 0.56883 & $(0.12806,\, 0.07345)$ \\
  & 1.00 & 2 & 4.98573  & 0.47653 & $(0.00000,\, 0.00000)$ \\
\midrule
4 & 0.00 & 2 & 11.28943 & 1.00000 & $(0.00000,\, 0.00000)$ \\
  & 0.25 & 1 & 11.16493 & 1.00000 & $(0.00194,\, 0.00021)$ \\
  & 0.50 & 1 & 10.60357 & 0.99128 & $(-0.00249,\, -0.00185)$ \\
  & 0.75 & 1 & 9.63979  & 0.93464 & $(0.00168,\, -0.00169)$ \\
  & 1.00 & 1 & 8.40737  & 0.83362 & $(0.00000,\, 0.00000)$ \\
\bottomrule
\end{tabular}
\end{table}

Frame-wise descriptors of the same trajectories complete the picture. The
Kozachenko--Leonenko estimator of equation~\eqref{eq:kl_entropy} returned
$\hat H(\rho_0) = -1.396695$, $-0.674706$, $0.258929$, and $0.938705$ nats at
the source endpoint of cases~1 through~4, and $\hat H(\rho_1) = -0.010401$,
$0.398166$, $0.244849$, and $-0.054806$ nats at the target endpoint. The
corresponding differences $\hat H(\rho_1) - \hat H(\rho_0)$ were $1.386294$,
$1.072872$, $-0.014080$, and $-0.993511$ nats, the value recorded for case~1
agreeing with $\log 4 = 1.386294$ to six decimal places. Both endpoint clouds
of cases~1 and~4 lie on parametric curves to which no perturbation noise was
applied, as recorded by the source and target noise of $0.0$ in the case~1
log, so for those two cases the endpoint estimates are formed on samples
whose support is one-dimensional and no interpretation as a two-dimensional
differential entropy is claimed here.

At the midpoint the estimator returned $\hat H(\rho_{1/2}) = 1.003174$,
$1.389257$, $1.195729$, and $1.864651$ nats, exceeding the corresponding
noise-only reference $H_{\mathrm{diff}}(1/2)$ of
equation~\eqref{eq:diffusive_reference}, which takes the values $-2.460440$,
$-1.544150$, $-2.054975$, and $-1.767293$ nats, by $3.463614$, $2.933407$,
$3.250704$, and $3.631944$ nats. The entropy attained an interior maximum in
every case, of $1.069491$, $1.423968$, $1.235989$, and $1.945279$ nats,
located at $t = 0.6303$, $0.4370$, $0.4790$, and $0.4228$ respectively. The
mean across frames of the ninety-five-percent half-width on $\hat H$ lay
between $0.029953$ and $0.033622$ nats in all four cases.

The RMS dispersion of equation~\eqref{eq:rms_dispersion} took the values
$1.000500$, $1.501205$, and $2.001001$ at $t = 0$, $1/2$, and $1$ in case~1,
increasing monotonically and departing from linearity at the midpoint by
$4.54 \times 10^{-4}$, which is below the mean half-width of $1.574 \times
10^{-3}$ for that case. In case~2 the dispersion increased monotonically from
$0.875659$ through $1.170203$ to $1.514369$. In case~3 it fell from
$0.998302$ to $0.940295$ at the midpoint before returning to $0.999436$, with
a recorded maximum of $1.003315$ at $t = 0.9832$. In case~4 it decreased
monotonically from $1.500751$ through $1.303508$ to $1.118593$. The mean
half-widths on the dispersion ranged from $1.574 \times 10^{-3}$ to $9.230
\times 10^{-3}$ across the four cases.

The eccentricity of equation~\eqref{eq:eccentricity} remained below the
resolution threshold $e = 0.25$ at every frame in cases~1 and~4, spanning
$[0.000000,\, 0.147554]$ and $[0.000064,\, 0.207380]$ respectively, so the
principal-axis angle is reported as undefined for those two cases and no
reorientation is quoted. In case~2 the eccentricity spanned $[0.042739,\,
0.471671]$ and the angle was resolved at $76$ of $120$ frames, giving first
and last resolved values of $-37.6281^\circ$ and $-59.1271^\circ$ and a net
folded reorientation of $-21.4990^\circ$, against a mean
ninety-five-percent half-width of $12.0446^\circ$. In case~3 the eccentricity
spanned $[0.141166,\, 0.883400]$ and the angle was resolved at $114$ of $120$
frames, giving first and last resolved values of $-18.6137^\circ$ and
$71.3167^\circ$ and a net folded reorientation of $89.9304^\circ$, against a
mean ninety-five-percent half-width of $2.8204^\circ$. The case~3
reorientation therefore departs from the imposed rotation of $90^\circ$ by
$0.0696^\circ$, a discrepancy some forty times smaller than the mean
half-width of the estimate, whereas the case~2 reorientation is smaller in
magnitude than twice its mean half-width. Table~\ref{tab:entropy} collects
the frame-wise summary.

\begin{figure}[H]
\centering
\includegraphics[width=\linewidth]{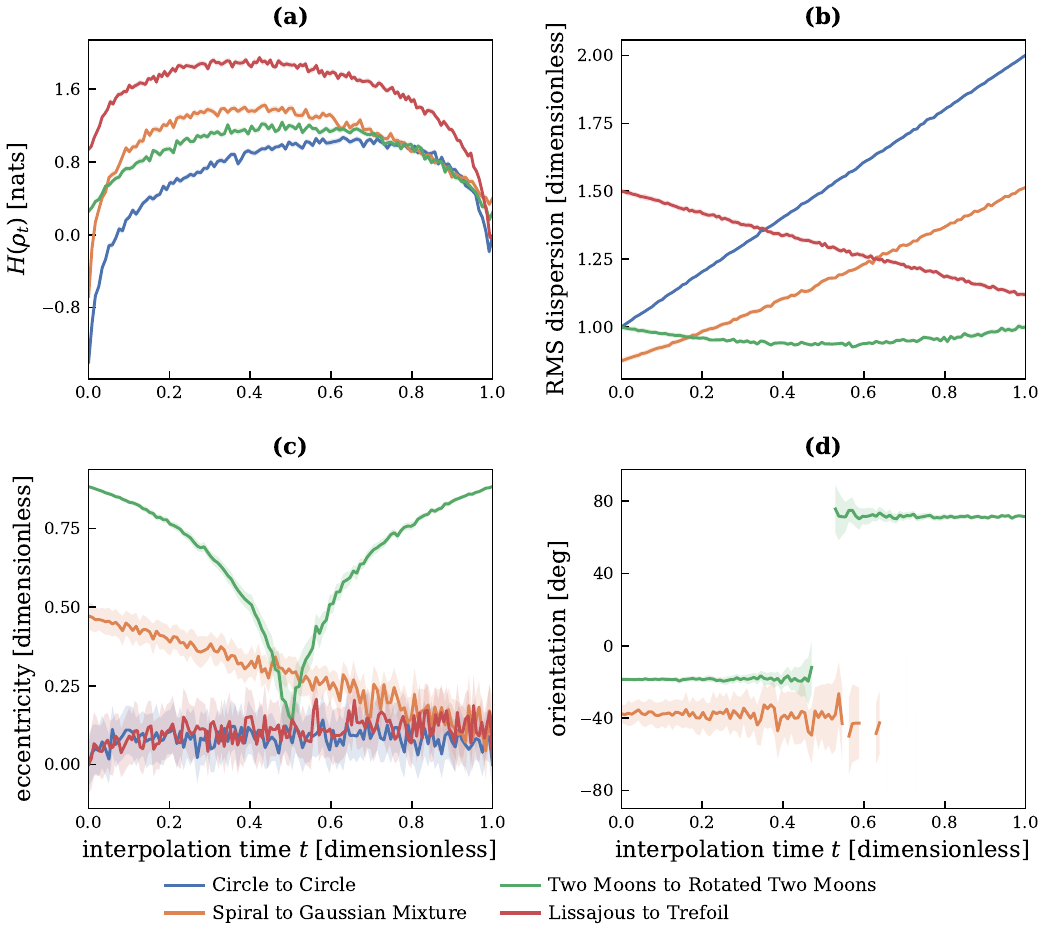}
\caption{Frame-wise descriptors of the bridge marginal against interpolation
time $t$. (a) Kozachenko--Leonenko differential entropy $\hat H(\rho_t)$ of
equation~\eqref{eq:kl_entropy} at $k = 5$. (b) RMS dispersion of
equation~\eqref{eq:rms_dispersion}. (c) Eccentricity of
equation~\eqref{eq:eccentricity}. (d) Principal-axis orientation, reported
only at frames whose eccentricity is at least $0.25$ and unwrapped within,
but never across, contiguous resolved runs. Shaded bands are the
ninety-five-percent intervals obtained by subsampling without replacement at
$B = 80$ replicates and $m/n = 0.8$, rescaled to full sample size through
equation~\eqref{eq:subsampling_rescale}. Cases~1 and~4 are absent from panel
(d), their eccentricity remaining below the resolution threshold at every
frame.}
\label{fig:entropy_production}
\end{figure}

Figure~\ref{fig:entropy_production} shows the four frame-wise descriptors
against interpolation time together with their subsampling uncertainty bands.
Figure~\ref{fig:entropy_production}(a) shows a concave profile in every case,
rising steeply away from $t = 0$ and falling steeply toward $t = 1$.
Figure~\ref{fig:entropy_production}(b) shows the monotone linear profile of
case~1, the monotone increase of case~2, the shallow interior minimum of
case~3, and the monotone decrease of case~4.
Figure~\ref{fig:entropy_production}(c) shows the low, flat eccentricity traces
of cases~1 and~4 alongside the pronounced interior minimum of case~3 at $t
\approx 0.5$. Figure~\ref{fig:entropy_production}(d) shows the resolved
orientation segments, the case~3 trace being interrupted over the interval on
which its eccentricity falls below threshold and resuming approximately
$90^\circ$ from its earlier level.

\begin{table}[H]
\centering
\caption{Frame-wise informational and geometric summary along the bridge.
Entropies are Kozachenko--Leonenko estimates of
equation~\eqref{eq:kl_entropy} in nats, evaluated at $k = 5$. The
reorientation is the folded net change in principal-axis angle, quoted with
the mean ninety-five-percent half-width across resolved frames; it is
undefined for cases~1 and~4, whose eccentricity remains below the resolution
threshold at every frame. The endpoint entropies of cases~1 and~4 are
evaluated on noiseless parametric curves.}
\label{tab:entropy}
\begin{tabular}{lccccccc}
\toprule
Case & $\varepsilon$ & $\hat H(\rho_0)$ & $\hat H(\rho_{1/2})$ & $\hat H(\rho_1)$ & $\max_t \hat H$ ($t$) & $\max_t \mathrm{rms}$ & Reorientation [deg] \\
\midrule
1 & 0.02 & $-1.3967$ & $1.0032$ & $-0.0104$ & 1.0695 (0.6303) & 2.0010 & --- \\
2 & 0.05 & $-0.6747$ & $1.3893$ & $0.3982$  & 1.4240 (0.4370) & 1.5144 & $-21.499 \pm 12.045$ \\
3 & 0.03 & $0.2589$  & $1.1957$ & $0.2448$  & 1.2360 (0.4790) & 1.0033 & $89.930 \pm 2.820$ \\
4 & 0.04 & $0.9387$  & $1.8647$ & $-0.0548$ & 1.9453 (0.4228) & 1.5008 & --- \\
\bottomrule
\end{tabular}
\end{table}

All wall-clock timings reported in this section were measured on a
single laptop, a Lenovo ThinkPad T440s (model 20AQ006HUS) with an
Intel Core i7-4600U processor (four logical cores at a maximum clock
of $3.30$~GHz), running Linux Lite~6.6 (x86\_64) under Linux kernel
5.15.0-185-generic, and therefore characterize the method on commodity
hardware rather than an optimized high-performance configuration. The run logs record the wall-clock cost of each case, decomposed into the
generation of the source and target clouds, solver initialization, the
Sinkhorn solve, trajectory generation, data serialization, and animation
rendering. The solve occupied $2.897$, $9.134$, $9.174$, and $8.217$ seconds
for cases~1 through~4, cloud generation and serialization together remained
below $0.4$ seconds in every case, and the total wall-clock time was
$61.341$, $54.383$, $59.965$, and $75.244$ seconds. Animation rendering
accounted for between $82.5\%$ and $91.8\%$ of the total in every case, so the
solver and its diagnostics constitute a minor fraction of the end-to-end
runtime. The single-sweep case~1 solve nevertheless required $2.897$ seconds,
larger than a per-sweep extrapolation from the other cases would predict,
because the just-in-time compilation of the solver kernels is performed once
per process and is charged to the first solve. Table~\ref{tab:timing}
collects the timing breakdown.

\begin{table}[H]
\centering
\caption{Wall-clock timing breakdown from the run logs, in seconds. The solve
column is the log-domain Sinkhorn iteration and includes the one-time
just-in-time compilation of the solver kernels; the visualization column is
the animation rendering. The thread count was left to automatic selection in
all runs.}
\label{tab:timing}
\begin{tabular}{lccccccc}
\toprule
Case & Distributions & Solver init & Solve & Trajectory & Serialization & Visualization & Total \\
\midrule
1 & 0.000 & 0.248 & 2.897 & 1.811 & 0.087 & 56.296 & 61.341 \\
2 & 0.001 & 0.005 & 9.134 & 0.087 & 0.292 & 44.863 & 54.383 \\
3 & 0.003 & 0.040 & 9.174 & 0.105 & 0.110 & 50.532 & 59.965 \\
4 & 0.000 & 0.005 & 8.217 & 0.106 & 0.076 & 66.839 & 75.244 \\
\bottomrule
\end{tabular}
\end{table}


\section{Discussion}

The four demonstration cases place the log-domain formulation in the regime
for which it was adopted. The ratio $\max_{ij} C_{ij} / \varepsilon$ lies
between $252$ and $400$ across the four cases, so the Gibbs kernel entries
$K_{ij} = \exp(-C_{ij}/\varepsilon)$ of~\eqref{eq:gibbs_kernel} span several
hundred decades and underflow the smallest representable double-precision
number over most of their range. An exponential-domain implementation of the
Sinkhorn recursion would therefore fail outright at these settings, whereas
the potentials $(f, g)$ of~\eqref{eq:log_potentials} remained bounded, the
marginal residual reached $10^{-9}$, and the reported marginal fidelity was
unity to eight decimal places in every case. This behavior is consistent with
the motivation for log-domain and stabilized scaling
schemes~\cite{schmitzer2019,peyrecuturi2019,chizat2018}.

The residual histories are geometric over approximately eight decades, in
qualitative agreement with the classical theory of the Sinkhorn recursion,
which establishes convergence and a geometric rate through the contraction
of the scaling map in the Hilbert projective
metric~\cite{sinkhorn1964,sinkhornknopp1967,franklin1989,knight2008}. The
quantitative content of that theory is nevertheless inaccessible at the
present parameters. The Birkhoff contraction ratio associated with the
kernel~\eqref{eq:gibbs_kernel} is $\kappa = (\sqrt{\eta} - 1) / (\sqrt{\eta}
+ 1)$ with $\eta = \exp(\max_{ij} C_{ij} / \varepsilon)$~\cite{franklin1989},
so that $1 - \kappa$ ranges from about $10^{-55}$ to $10^{-87}$ over the
four cases. The guaranteed rate is thus numerically indistinguishable from
unity and predicts no useful decay, while the observed per-iteration factors
lie between $0.966$ and $0.976$. The measured decay is therefore an
asymptotic local rate attained near the fixed point rather than a
realization of the worst-case bound, and the gap between the two is many
orders of magnitude. Consistent with this, the three fitted factors do not
order with $\varepsilon$. Because the present design varies $\varepsilon$
and the source and target geometry together, and because only three cases
admit a fit, the data cannot separate the two influences, and no relation
between the contraction rate and the regularization parameter is claimed
here. A sweep in $\varepsilon$ at fixed geometry would be required to
address that question and is left to future work.

Case~1 warrants separate comment, since its convergence at the first
residual evaluation is a structural rather than a numerical result. The
source and target of~\eqref{eq:case1_source} are both sampled at angles that
are equispaced up to a global phase, so the squared-distance cost satisfies
$C_{ij} = r_0^2 + r_1^2 - 2 r_0 r_1 \cos( 2\pi (i - j) / n + \Delta\varphi )$
and depends on the indices only through $(i - j) \bmod n$. The cost matrix is
consequently circulant, and for a circulant cost with uniform marginals the
dual potentials that solve~\eqref{eq:eot_discrete} are constant vectors,
which the updates~\eqref{eq:f_update}--\eqref{eq:g_update} reach in a single
sweep from the zero initialization. The recorded residual of $3.771506 \times
10^{-15}$ is the floating-point signature of an exactly attained solution
rather than of a rapidly converging iteration. Two independent diagnostics
corroborate the interpretation: the row entropies of the conditional
coupling have a standard deviation of $4.15 \times 10^{-4}$ nats across the
retained block, every row being a cyclic shift of every other, and the
barycentric displacement has mean and maximum differing by $4.3 \times
10^{-5}$. Case~1 accordingly verifies that the implementation recovers the
closed-form solution where one exists, but it does not exercise the
iteration, and it should not be read as a convergence baseline.

The conditional structure of the couplings behaves broadly as the
regularization would suggest, the most diffuse coupling occurring at the
largest $\varepsilon$ and the sharpest at the smallest. This ordering is
carried by the effective sparsity $\exp(H(\pi^\star))/(nm)$, which rises from
$0.046568$ at $\varepsilon = 0.02$ to $0.110043$ at $\varepsilon = 0.05$, and
which measures the joint perplexity of the plan as a fraction of the $nm$
available cells. The ordering is not strict at the level of the conditional
perplexity, since case~3 at $\varepsilon = 0.03$ carries more effective
targets per source than case~4 at $\varepsilon = 0.04$: the diffuseness of a
row is set by $\varepsilon$ relative to the local separation of target points
rather than by $\varepsilon$ alone, and the four geometries differ in that
separation. The transport cost recorded in the logs is likewise geometric in
origin, being largest for case~1, whose supports are separated by a full unit
of radius, and smallest for case~4, whose interleaved lobes require the least
displacement.

Interpretation of the reported conditional quantities requires care on a
separate count. The plan is archived on a strided $500 \times 500$ block of a
$1000 \times 1000$ coupling, which retains one quarter of the entries. If the
retained entries are representative of the whole, the renormalized block
entropy satisfies $H_{\mathrm{block}} = H(\pi^\star) - \log 4$, and the
differences reported above agree with $\log 4$ to within $2.8 \times 10^{-3}$
nats in all four cases. That agreement is an internal consistency check that
the strided block does carry a representative quarter of the mass. The same
argument applied at the level of a single row, where one half of the targets
is retained, gives $H_i^{\mathrm{block}} = H_i^{\mathrm{full}} - \log 2$ and
hence a perplexity that is understated by a factor of two. The reported
values of $23.2840$, $58.0584$, $40.7015$, and $28.9141$ therefore correspond
to full-plan effective target counts of approximately $46.6$, $116.1$,
$81.3$, and $57.8$, and the reported peak conditional probabilities
correspondingly overstate the full-plan values by close to a factor of two.
The barycentric displacements are unaffected, the conditional mean over a
uniformly strided subset of a smooth conditional being unbiased. Archiving
the full plan, at a cost of $8 n^2$ bytes, would remove the need for this
correction and is the preferable course for point-cloud sizes at which the
storage remains tractable.

The mean barycentric displacement of case~1, $0.994994$, falls short of the
exact radial separation of unity by $0.5\%$. The shortfall is not a
numerical defect but a property of the barycentric projection at finite
regularization. The conditional expectation~\eqref{eq:barycentric_map}
averages the target points over an arc whose angular width is set by
$\varepsilon$, and the average of points on a circle over an arc of nonzero
width lies strictly inside that circle. The image radius is therefore smaller
than $2$, and the deficit contracts as $\varepsilon \to 0^+$, in which limit
the coupling concentrates on the graph of the Brenier map and the barycentric
projection recovers it exactly~\cite{brenier1991,mikami2004,carlier2017}. The
same mechanism underlies the difference between the barycentric map and the
sampled bridge of~\eqref{eq:bridge_sample}, the former reporting a
conditional mean and the latter a draw from the conditional itself.

The entropy estimates require the most careful reading of any quantity
reported here, and the endpoint values of cases~1 and~4 do not admit the
interpretation that their name suggests. The Kozachenko--Leonenko estimator
and its analysis presuppose a distribution that is absolutely continuous
with respect to Lebesgue measure on $\mathbb{R}^d$, a hypothesis under which
the bias and variance have been characterized in
detail~\cite{kozachenko1987,delattre2017,berrett2019}. The source and target
clouds of cases~1 and~4 carry no perturbation noise, as the source and target
noise of $0.0$ recorded in the case~1 log confirms, and they lie exactly on
one-dimensional parametric curves, which are Lebesgue-null in the plane. The
corresponding differential entropy is not defined, and the finite values
returned by~\eqref{eq:kl_entropy} are artifacts of finite sample size. The
mechanism is explicit in the estimator: for $n$ points spread along a
rectifiable curve of length $L$, the $k$-th nearest-neighbor distances scale
as $r_i \sim L / n$, so that the sum in~\eqref{eq:kl_entropy} contributes
$2 \log(L/n)$ while the digamma term contributes $\log n$, giving $\hat
H(\rho) \approx \mathrm{const} + 2 \log L - \log n$. Two consequences follow
and both are visible in the reported numbers. First, the estimate diverges
logarithmically as the sample grows, so the recorded value of $\hat H(\rho_0)
= -1.396695$ for case~1 is a statement about $n = 1000$ and not about the
source distribution. Second, differencing two curve-supported estimates at
equal $n$ cancels the sample-size term and isolates $2 \log(L_1 / L_0)$. For
case~1 both endpoints are circles of radii $1$ and $2$, so the prediction is
$2 \log 2 = 1.386294$ nats, which is precisely the recorded value of $\hat
H(\rho_1) - \hat H(\rho_0)$ to six decimal places. The quantity tabulated as
the net entropy production of case~1 is therefore a measurement of the ratio
of two curve lengths. The corresponding value for case~4, $-0.993511$ nats,
carries the same character and reflects the trefoil projection being shorter
than the Lissajous curve of equal parametric sampling. Case~3 provides the
internal control that confirms the diagnosis, its endpoints carrying additive
noise of standard deviation $0.05$ and therefore being genuinely
two-dimensional: its recorded difference of $-0.014080$ nats is consistent
with zero, as the isometry relating its two endpoint distributions requires.
Case~2, whose source carries only a small perturbation of standard deviation
$0.01$ while its target is a genuine mixture of Gaussians, occupies an
intermediate position and its endpoint difference should not be interpreted
quantitatively either. The conservative reading of the present data is that
$\hat H(\rho_t)$ is informative on the open interval $t \in (0, 1)$, where
the bridge noise of variance $\varepsilon\, t(1-t)$ renders every marginal
absolutely continuous, and that the two endpoint columns should be excluded
from any comparison. Applying a small perturbation to the case~1 and case~4
generators would place all four cases on a common footing and is the natural
remedy.

Read on the open interval, the entropy profiles are consistent with the
structure of the bridge. Every case attains an interior maximum, which
follows from the bridge variance $\varepsilon\, t(1-t)$
of~\eqref{eq:bridge_marginal} vanishing at both ends and peaking at $t =
1/2$~\cite{follmer1988}. The measured midpoint entropies exceed the
noise-only reference~\eqref{eq:diffusive_reference} by between $2.93$ and
$3.63$ nats, confirming that the marginal at the midpoint is not dominated by
the diffusive term and retains substantial geometric structure inherited from
the endpoints. The maxima are displaced from $t = 1/2$ in a direction that
tracks the deterministic part of the interpolation: case~3, whose endpoints
are related by an isometry, peaks at $t = 0.4790$, closest to the symmetric
value, whereas case~1, whose support lengthens monotonically, peaks late at
$t = 0.6303$ and case~4, whose support shortens, peaks early at $t = 0.4228$.
The symmetric noise contribution alone would place every maximum at $t =
1/2$, so the displacement measures the asymmetry of the drift term.

The geometric descriptors supply the principal quantitative verification of
the pipeline. The target of case~3 is generated by rotating a two-moons cloud
through $\pi/2$, and the recovered net reorientation of the covariance
principal axis is $89.9304^\circ$, departing from the imposed value by
$0.0696^\circ$, roughly forty times smaller than the mean
ninety-five-percent half-width of $2.8204^\circ$. The recovery is achieved
across a masked interval near $t = 1/2$ over which the cloud passes through
near-isotropy, the eccentricity falling to $0.141166$, and over which the
principal axis is genuinely unobservable. Declining to unwrap the angle
across that gap costs nothing here, since the two resolved segments differ by
approximately the imposed rotation, and it avoids attributing to the data a
branch choice the data do not determine~\cite{mardia2000}. The contrast with
case~2 is instructive: its net reorientation of $-21.4990^\circ$ is smaller
in magnitude than twice its mean half-width of $12.0446^\circ$ and is
resolved at only $76$ of $120$ frames, so the present data do not support a
claim of net axis reorientation in that case. Cases~1 and~4 remain below the
eccentricity threshold at every frame, which is the expected behavior for a
circle and for two curves possessing rotational symmetry of order greater
than two, whose covariance is isotropic by construction.

The timing breakdown recorded in the logs situates the computational cost of
the method. The Sinkhorn solve occupied between $2.897$ and $9.174$ seconds
per case, and the generation of the distributions, the trajectory, and the
serialized archive together remained below one second in every case, so the
transport computation and its diagnostics are a minor part of the end-to-end
runtime. Animation rendering dominated the total, accounting for between
$82.5\%$ and $91.8\%$ of the wall-clock time, which reflects a design choice
to emit a rendered interpolation film rather than a property of the solver.
The single-sweep case~1 nevertheless spent $2.897$ seconds in the solve,
because the just-in-time compilation of the numerical
kernels~\cite{lam2015} is performed once per process and charged to the
first solve; the compiled cost per sweep, inferred from the multi-hundred
sweep cases, is on the order of ten milliseconds. Because the thread count
was left to automatic selection in every run and the bridge sampler draws its
randomness within a thread-parallel region, the timings and the sampled
trajectories are tied to the thread configuration of the host, and bitwise
reproducibility across differing thread counts is not guaranteed, in common
with the associative reordering permitted by the compiler options adopted
here~\cite{goldberg1991}.

Three features of the analysis limit the strength of the conclusions and are
stated here rather than left to inference. First, the case~3 target cloud has
its centroid at the origin while the source cloud has its centroid at
$(0.50166, 0.25285)$, so the rigid motion realized between the two clouds is a
rotation composed with a translation of magnitude $0.56178$ rather than a
rotation about a common centroid. The transport cost of $0.866267$ and the
barycentric displacements of case~3 accordingly contain a translational
contribution and are not a pure measure of angular reorganization. The
covariance-based descriptors are translation-invariant and are unaffected, so
the recovery of the rotation angle discussed above is not compromised.
Second, the three interior snapshot times of the density montage do not
coincide with stored frames on either the $120$-frame or the $150$-frame grid
and are obtained by linear interpolation between straddling frames. Because
consecutive frames are independent draws from their respective marginals
rather than points of a common Lagrangian path, a weighted blend of two such
frames with weights $w$ and $1 - w$ carries bridge noise of variance reduced
by a factor $w^2 + (1 - w)^2$, equal to $0.5$ at $t = 1/2$ and $0.625$ at $t
= 0.25$ and $t = 0.75$, and additionally replaces each particle's target by a
convex combination of two independently drawn targets. The interior panels
therefore understate the dispersion of $\rho_t$ and are contracted slightly
toward the barycentric image. Choosing $N_f$ so that the snapshot times fall
on stored frames, for which $N_f = 121$ suffices, would remove the artifact.
Third, the region count $N_{\mathrm{reg}}(t)$ of~\eqref{eq:region_count} is a
count of connected components of a fixed super-level set of a KDE
and is therefore contingent on both the level and the
bandwidth~\cite{scott1979,silverman1986}, and it is not a topological
invariant of the underlying support. Its value is interpretable where the
target possesses well-separated components, as in the sequence $1, 3, 4, 4,
4$ of case~2, which tracks the fragmentation of a connected curve into the
four components of the mixture. It is less informative for the ridge-like
supports of cases~1 and~4. The support fraction is referred to a per-case
bounding box, saturates at unity for case~4 at the first two snapshots, and
should not be compared across cases; the effective area, which carries units
of area through the cell-area factor in~\eqref{eq:effective_area}, is the
more robust of the two coverage measures.

Several further limitations bound the scope of the present study. All four
cases fix $n = m = 1000$ and $d = 2$, so neither the scaling of the solver
with point-cloud size nor its behavior in higher dimensions is assessed here;
the $O(nm)$ cost per sweep and the $8nm$-byte footprint of the dense cost
matrix are the binding constraints on the former. The regularization is fixed
within each case, so the entropic bias of the coupling is not resolved as a
function of $\varepsilon$. The diagnostics are computed on single
realizations at a fixed base seed, and the uncertainty bands quantify
sampling variability within a realization rather than variability across
independent runs. Finally, the four cases are constructed rather than
measured, and the extent to which the observed behavior transfers to
empirical point clouds arising in applications, where the marginals are noisy
and possibly of unequal mass, remains to be established. The unbalanced and
multi-marginal extensions of the scaling
framework~\cite{chizat2018,benamou2015} provide the natural setting in which
those questions would be posed.

\section{Conclusions}

This work presented \texttt{anyakrakusuma}, a Python library that solves the
discrete static Schr\"{o}dinger bridge problem through a log-domain
Sinkhorn--Knopp iteration and reconstructs the entropic interpolation between
two empirical point clouds, together with a diagnostic pipeline exercised on
four idealized planar cases spanning a circle-to-circle dilation, a
spiral-to-mixture fragmentation, a rigid reorientation of two moons, and a
Lissajous-to-trefoil deformation. The log-domain formulation was necessary
rather than merely convenient at the parameters studied: the
cost-to-regularization ratio reached four hundred, at which the Gibbs kernel
underflows double precision across most of its range, yet the iteration
reached a marginal residual of $10^{-9}$ and unit marginal fidelity in every
case, with geometric residual decay over approximately eight decades at
per-iteration contraction factors between $0.966$ and $0.976$. These are
local rates attained near the fixed point and lie many orders of magnitude
below the worst-case Hilbert-metric bound, which is vacuous at these settings;
the circle-to-circle case converged in a single sweep because its equispaced
angular sampling renders the cost matrix circulant, a structural exactness
that serves as a correctness check rather than a convergence baseline. The
covariance analysis recovered the imposed ninety-degree reorientation of the
two-moons case to within $0.07^\circ$, roughly forty times smaller than the
estimator's uncertainty and across a masked interval of near-isotropy on which
the principal axis is unobservable, which is the strongest quantitative
validation the pipeline provides.

The analysis equally delimited what the diagnostics cannot support, and these
boundaries are as much a part of the result as the recoveries: the
differential entropy is well defined only on the open interpolation interval,
the endpoint estimates of the two noiseless cases measure curve length rather
than entropy, the conditional coupling statistics carry an exact
factor-of-two offset from the subsampled plan storage, the rotation case
realizes an unintended rigid translation, and the interior density snapshots
are temporally interpolated between independently sampled frames, each with a
stated remedy in perturbing the noiseless generators, archiving the full plan,
recentering the rotated target, and aligning the frame count with the snapshot
times. Future work follows directly from these limitations: a regularization
sweep at fixed geometry would separate the influence of $\varepsilon$ on the
contraction rate from that of the transport geometry, while systematic study
of the solver's scaling with sample size and dimension, replacement of the
dense cost matrix by stabilized sparse scaling for larger
problems~\cite{schmitzer2019}, and extension to unbalanced and multi-marginal
settings~\cite{chizat2018,benamou2015} would move the library from idealized
demonstrations toward the empirical point clouds that motivate entropic
transport in practice, with the software, configurations, and diagnostic
scripts released openly so that the present results can be reproduced and
extended.

\section*{Acknowledgements}
The authors used Claude Sonnet~5 (Anthropic, PBC) solely as a
writing-assistance tool to refine English vocabulary and grammar
during the preparation of this manuscript. All scientific content,
interpretations, analyses, conclusions, and any remaining linguistic
imperfections are the sole responsibility of the authors.

\section*{Funding}
This study was funded by the Indonesian Ministry of Education,
Culture, Research, and Technology 2026
(169/C3/DT.05.00/PL-BARU/2026).

\section*{Author Contributions}
\textbf{D.E.I.}: Conceptualization, Data curation, Formal analysis,
Investigation, Funding acquisition,
Project administration, Supervision, Resources, Writing -- original draft.
\textbf{S.H.S.H.}: Conceptualization, Data curation, Formal analysis,
Investigation, Methodology, Software, Visualization, Validation,
Writing -- original draft. \textbf{A.W.J.}: Supervision, Writing -- review and editing.
\textbf{S.F.B.}: Supervision, Writing -- review and editing.
\textbf{C.S.D.}: Supervision, Writing -- review and editing.
\textbf{E.R.}: Supervision, Writing -- review and editing.
\textbf{A.P.}: Supervision, Writing -- review and editing.
\textbf{R.D.K.}: Supervision, Writing -- review and editing.
\textbf{R.S.}: Supervision, Resources, Writing -- review and editing.
\textbf{D.J.P.}: Supervision, Resources, Writing -- review and
editing.

\section*{Data Availability}
The \texttt{anyakrakusuma} library source code is available on GitHub at \url{https://github.com/sandyherho/anyakrakusuma} and from the Python Package Index at \url{https://pypi.org/project/anyakrakusuma/}. The supplementary data-analysis scripts that reproduce the diagnostic metrics and figures are available at \url{https://github.com/sandyherho/suppl_anyakrakusuma}. All supplementary outputs, comprising the raw NetCDF archives, the computed diagnostic metrics, the run logs and timing breakdowns, and all figures, are permanently archived on the Open Science Framework at \url{https://doi.org/10.17605/OSF.IO/VQWF4}. The library, the supplementary scripts, and the archived outputs are all released under the MIT license.

\end{document}